%% file: main.tex
\documentclass[pra, twocolumn,
 superscriptaddress,
 amsmath,amssymb,
 aps,
floatfix, 
longbibliography
]{revtex4-1}
\usepackage{graphicx}
\usepackage{bm}

\usepackage{xcolor}
\usepackage[colorlinks=true,citecolor=blue,linkcolor=magenta]{hyperref}
\usepackage{orcidlink}

\usepackage[latin9]{inputenc}
\usepackage{amsmath}
\usepackage{amssymb}
\usepackage{graphicx}
\usepackage{verbatim}
\usepackage{bm}
\usepackage{color}

\usepackage{graphicx,epsfig,amsfonts,amssymb}
\usepackage{bm}
\usepackage{times}
\usepackage{lipsum}
\usepackage{verbatim}

\newcommand{\la}{\langle}
\newcommand{\ra}{\rangle}

\newcommand{\be}{\begin{eqnarray}}
\newcommand{\ee}{\end{eqnarray}}

\newcommand{\bs}{\begin{equation}\begin{split}}
\newcommand{\es}{\end{split}\end{equation}}

\input{shortcut.tex}

\date{\today}

\begin{document}

\title{Competing Bosonic Reactions: Insight from Exactly Solvable Time-Dependent Models}

\author{Fumika Suzuki\orcidlink{0000-0003-4982-5970}}\email{fsuzuki@lanl.gov}
\affiliation{Theoretical Division, Los Alamos National Laboratory, Los Alamos, New Mexico 87545, USA}
\affiliation{Center for Nonlinear Studies, Los Alamos National Laboratory, Los Alamos, New Mexico 87545, USA}
\author{Rajesh K. Malla\orcidlink{0000-0003-1052-3705}}
\affiliation{Physical Review Letters, American Physical Society, Hauppauge, New York 11788, USA}
\author{Nikolai A. Sinitsyn\orcidlink{0000-0002-0746-0400}}\email{nsinitsyn@lanl.gov}
\affiliation{Theoretical Division, Los Alamos National Laboratory, Los Alamos, New Mexico 87545, USA}

\begin{abstract}

   
   We discuss the progress on exactly solvable multistate Landau-Zener models from a perspective of their application to competing reactions of particle creation from a false vacuum. Such models generally predict that, even with identical initial conditions, and for nearly the same other particle parameters, a quantum coherent evolution  results in a  final particle distribution with significant asymmetry. We use an exact solution of the driven bosonic Tavis-Cummings model for two reaction pathways in order to quantify this effect, reveal a corresponding phase transition, and identify its universality class.
\end{abstract}
\maketitle
\section{Introduction}
Multistate Landau-Zener (MLZ) models describe nonadiabatic dynamics in a wide range of driven quantum systems that evolve according to the time-dependent Schr\"odinger equation
\begin{equation}
i\frac{\partial}{\partial t} |\Psi(t) \ra = H(t) | \Psi (t) \ra,
    \label{se}
\end{equation}
where $H(t)$ is the matrix Hamiltonian that has either constant or  linear in time, $t$, parameters:
\begin{equation}
  H(t) =Bt+A,
    \label{Hmlz}
\end{equation}
and where $B$ and $A$ are Hermitian time-independent matrices \cite{Brundobler1993}. The MLZ models arise in the physics of Rydberg atoms \cite{Harmin1994,Harmin1997}, quantum dots \cite{Shevchenko2010,Petta2018,Burkard2020},  and circuit QED systems \cite{Werther2019, Bello2020, Kervinen2019,Sun2012,Wen2020, Malla2018}. 
In general, they are not solvable but, for special cases, many fully solvable instances have been recently identified  \cite{Sinitsyn2018,YUZBASHYAN2018323,Sinitsyn2016,Chernyak2020,Malla2021,Barik2024}.  By ``solvable" we mean here that for all possible values of  nonzero elements of $A$ and $B$ in a given model we can write the elements of the scattering matrix, for the evolution of the state vector during $t\in (-\infty, +\infty)$, in terms of the conventionally known special functions of these parameters.


The {\it driven bosonic Tavis-Cummings model}, which we will define precisely later in Eq.~(\ref{h23}),  is the most complex MLZ model that has been solved to date \cite{Malla2022,Sinitsyn2016,Sun2016,Sun2019}. Although the Tavis-Cummings model was originally developed to describe two-level atoms interacting with a quantized bosonic field, it has been shown that the model can be mapped onto a Hamiltonian describing a transition of a multi-particle  bosonic condensate through a resonance, during which the initial bosons decay into pairs of new bosonic particles. 
This process has a natural interpretation in terms of a molecular condensate dissociating into individual atoms via a coherent reaction of the form 
\be
ab \leftrightarrow a+b,
\label{reac}
\ee
in which a bosonic molecule, $ab$, can dissociate reversibly into a pair of new bosons, $a$ and $b$. 

In a field theory Lagrangian commonly used in cosmology and high-energy physics, the analogous process is represented by a cubic interaction potential 
\begin{equation}
V_{int} \sim g \Psi \phi^2, 
\label{cube}
\end{equation}
where $\Psi$ is the original scalar field. 
The system  can be initially in the pseudo-vacuum state with
a macroscopic number of bosons occupying the lowest energy level,
and $\phi$ is the field of the emerging new particles. 
Taking into account the analogy between the process described by the driven bosonic Tavis-Cummings model  and that described by a field theory Lagrangian, we can interpret $\Psi$ in (\ref{cube}) as the molecular field, and  $\phi$ as to the field of  atoms. We will assume that the product particles differ by their internal degrees of freedom, e.g., momenta or spin directions. Such many-body coherent reactions have already been realized in experiments with ultra-cold atoms \cite{Zhang2021}. We note that such processes are also encountered in cosmology, e.g., the models of inflationary universe \cite{Lozanov2020}.   It has also been demonstrated that the Kibble-Zurek mechanism, originally formulated in the context of cosmology, exhibits an analogy to LZ models \cite{kibble,zurek,damski}.

For a coherent evolution in a  confined isolated volume, the interactions (\ref{cube}) do not ensure a high   reaction efficiency. To dissociate a significant portion of the molecules, the chemical potentials of the particles before and after the reaction must be  positioned at the resonance, a condition that is difficult to achieve with precision. Nonetheless,  the interacting system can be made to {\it pass through} such a resonance when subjected to time-dependent fields that vary the relative chemical potentials of the reacting bosons. Close to the resonance, the time-dependence of the bosonic chemical potentials can be approximated linearly, leading to a time-dependent mass term in the Lagrangian \cite{Tyagi2025}:
$$
V_{m}=\beta t  \Psi^2 +\frac{m  \phi_a^2}{2}+\frac{m \phi_b^2}{2}, 
$$
where $\beta$ is the rate of the transition through the resonance and $m$ is the time-independent mass of the emerging particles.

Such quantum mechanical dissipationless time-dependent  models are found in cosmology. For example, similar cubic interactions and linearly time-dependent background fields are encountered in physics of axions - the hypothetical particles whose interactions  (\ref{cube}) can describe the decay of an axion into pairs of gluons \cite{axion}.  

The emergence of matter-antimatter asymmetry in our universe requires   $CP$-symmetry violation in high energy physics, which has been found  in the decay rates of kaons, $K$ and $\bar{K}$ \cite{CP1}. However, the mass difference between these bosons is very small in comparison to the other possibly relevant interaction scales, such as the mass of the Higgs boson. Hence, even if there were a hypothetical condensate of bosons that had become unstable at some cosmological stage and  decayed into pairs of $K$ and $\bar{K}$ bosons, the coupling $g$ in (\ref{cube}) for this process was much larger than the known difference of $K$ and $\bar{K}$ masses. Then, how such a small asymmetry could lead to a considerable effect on the final state after passing through the resonance?

The goal of this article is to attract attention to an unusual property of   recently solved MLZ models, demonstrating that the result of the time evolution with the Hamiltonian (\ref{Hmlz}) can depend  on a small constant parameter strongly.  In particular, the complete integrability of the driven Tavis-Cummings model extends to multichannel reactions, in which the original field $\Psi$ can decay via different reaction channels, e.g., 
\be
ab \leftrightarrow a_1+b_1, \quad {\rm and } \quad ab \leftrightarrow a_2+b_2,
\label{reac2}
\ee
where indices $1$ and $2$ denote potentially different reaction products.
Thus, we can study a competition between  different reactions. The sensitivity of dynamics near a phase transition involving small symmetry breaking interactions was demonstrated in 
spin chains and well-mixed mean-field models \cite{Yan2021,Tyagi2025}. The bosonic Tavis-Cummings model can be used to illustrate this sensitivity without further approximations.  
Many results on this model  have  already been presented. Our goal here is to provide a more detailed and unified introduction to the driven Tavis-Cummings model, with the  amplification of a small asymmetry during the particle production as a unifying theme.  

The structure of this article is as follows. In section~\ref{3DO-sec}, we discuss the simplest solvable 3-state Landau-Zener (LZ) model that demonstrates the effect of large sensitivity of the scattering probabilities to a small parameter. 
This elementary system already exhibits many properties that we will discuss for truly many-body systems in the following sections. Sections~\ref{integ-MLZ} and~\ref{single-ch}  review previously established facts about the MLZ integrability and  the driven Tavis-Cummings model, with only minor refinements of the formulas, which are needed for the following discussion. 

 Section~\ref{TC-sed} is central to our article. It introduces a model of competition between different reaction channels and derives scaling for different types of excitations from the exact solution. In section~\ref{PT}, we then analyze physics of this process using time-scale separation between the processes of the original vacuum decay and the subsequent pseudo-thermalization of the product particles. In particular, we present a novel solvable model that describes this thermalization and proves the presence of a second phase transition responsible for the power law scaling of the excitations. Finally, in section~\ref{semicl}, we explore the semiclassical limit of the model with reaction competition, and reveal its relation to the universality class of the second-order phase transition with a broken symmetry that was introduced in \cite{Tyagi2025}.

\section{Three-state Demkov-Osherov model}
\label{3DO-sec}

 For MLZ models,  it is convenient to work in the {\it diabatic basis} of time-independent states that diagonalize $B$ in (\ref{Hmlz}). If then some of the elements, $B_{kk}$ and $B_{jj}$, of $B$ are degenerate, the corresponding orthogonal diabatic states, $|k\ra$ and $|j\ra$, are chosen so that they are not directly coupled; that is, in the diabatic basis 
$$
{\rm if}\quad  B_{kk}=B_{jj}, \quad {\rm then}\quad A_{kj}=0.
$$
{\it Diabatic energies} are the diagonal elements of the Hamiltonian (\ref{Hmlz}) in the diabatic basis. In a time-energy diagram, they are straight lines, $B_{kk}t+A_{kk}$, called {\it diabatic levels}. The off-diagonal elements of $A$ in the diabatic basis are referred to as {\it couplings} between the diabatic levels.

\begin{figure}[t!]
\centering \includegraphics[width=0.85\columnwidth]{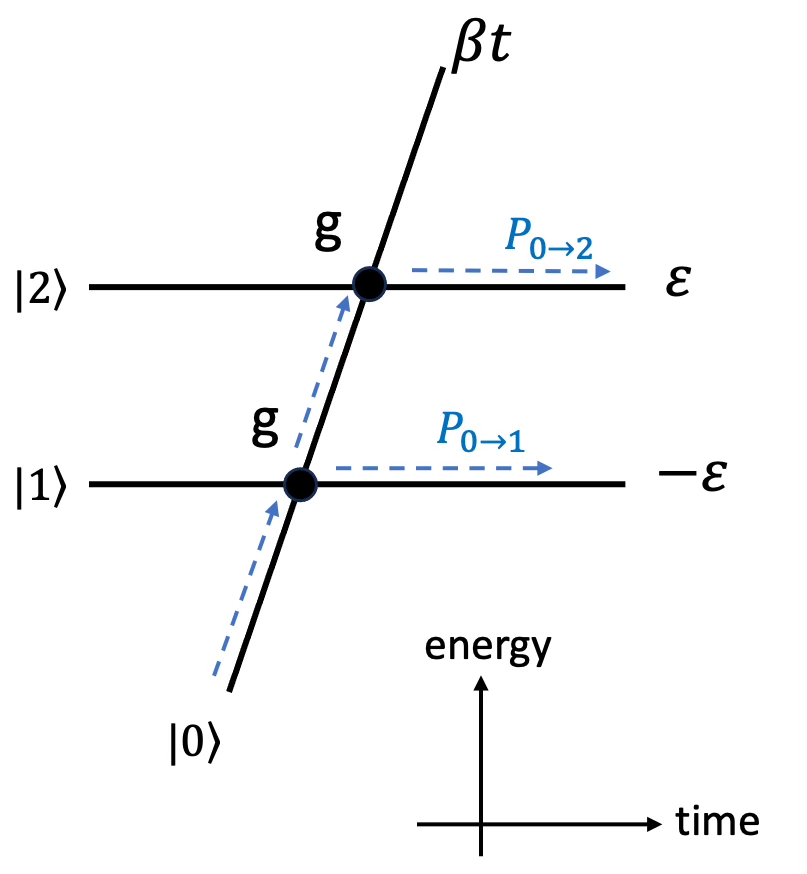}
\caption{Diabatic energy levels of the 3-state Demkov-Osherov model (\ref{do-3}). The corresponding diabatic states are  $|0\ra,\, |1\ra,\, |2\ra$. Black circles mark the directly interacting level pairs. Dashed arrows indicate the directions of transitions with probabilities $P_{0\rightarrow 1}$ and $P_{0\rightarrow 2}$.
}
\label{DO-fig}
\end{figure}
Note that as $t\rightarrow \pm \infty$ the diabatic states coincide with the instantaneous eigenstates of $H(t)$ in (\ref{Hmlz}). Therefore, the transitions between the diabatic levels at the infinite time are suppressed. One of the  goals of the MLZ theory is to find the scattering probabilities between any pair of diabatic states as time changes from $-\infty$ to $+\infty$. Namely, let as $t\rightarrow -\infty$ the system be in $j$-th diabatic state.  The transition to level $k$ probability  is then  the probability of finding the MLZ system in the $k$-th diabatic state as $t \rightarrow +\infty$.

\vspace{0.2cm}
\paragraph*{\it Three-state model.} An example of a MLZ Hamiltonian, written in the diabatic basis is
\begin{equation}
   H(t) =\left(
   \begin{array}{ccc}
    \beta t & g &g \\
    g & -\varepsilon &0  \\
    g & 0  & \varepsilon
   \end{array}
   \right), \quad \beta,\, \varepsilon>0,
    \label{do-3}
\end{equation}
where $g$, $\varepsilon$, and $\beta$ are constants. The corresponding diabatic levels are shown in Fig.~\ref{DO-fig}.

Let us assign the level with the finite slope an index $0$, while the other levels will have indices increasing with their diabatic energies. We will denote the corresponding diabatic states as $|0\ra$, $|1\ra$, and $|2\ra$. The model (\ref{do-3}) is a special case of the fully solvable Demkov-Osherov model \cite{Demkov1967, Sinitsyn2017}, which describes the interactions of a single level with a band of parallel energy levels. In particular, the exactly found transition probabilities, for the evolution starting at level $0$ as $t\rightarrow - \infty$ and ending as $t\rightarrow +\infty$, are known analytically exactly (see, e.g., in \cite{Sinitsyn2017} or the derivation later in this article):  
\begin{eqnarray}
\nonumber P_{0\rightarrow 0}& =& e^{-4\pi g^2/\beta}, \quad P_{0\rightarrow 1} =1- e^{-2\pi g^2/\beta}, \\
P_{0\rightarrow 2} &=&e^{-2\pi g^2/\beta} \left( 1- e^{-2\pi g^2/\beta}\right). 
\label{DO3P}
\end{eqnarray}

\vspace{0.2cm}
\paragraph*{\it Invariance of transition probabilities.} An unexpected property of Eq.~(\ref{DO3P}) is that the transition probabilities  do not depend on the diabatic energies $ \pm \varepsilon$.  As $\varepsilon \rightarrow 0$, the levels $1$ and $2$ become indistinguishable. However, even then
the effect of this parameter does not disappear because $P_{0\rightarrow 1} \ne P_{0\rightarrow 2}$. It is also useful to note a similarity of the final state probabilities and the probabilities of microstates in the equilibrium Gibbs distribution:
\be
\frac{P_{0\rightarrow 2}}{P_{0\rightarrow 1}} = e^{-2\pi g^2/\beta} =e^{-2\varepsilon/k_BT }, 
\label{ratio1}
\ee
where $2\varepsilon$ is the energy difference between the two parallel levels and 
\be
k_B T =\varepsilon \beta /(2\pi g^2)
\label{teff1}
\ee
is an effective temperature that depends on the  sweep rate $\beta$, the coupling $g$, and the level splitting itself. The latter fact is what makes this temperature sufficiently small in the limit $\varepsilon \rightarrow 0$ to guarantee a finite population difference.

This result may seem counterintuitive, given that apart from $\varepsilon$ the energy combinations that we can construct from the remaining parameters, $g$ and $\beta$, are $g$, $\beta^{1/2}$, and $\beta/g$. They can all be arbitrarily larger than $\varepsilon$, yet still lead to the asymmetry (\ref{ratio1}). 
Nevertheless, the role of the parameter $\varepsilon$ in this asymmetry is crucial. For example, if we set $\varepsilon=0$ in Eq.~(\ref{se}) with the Hamiltonian (\ref{do-3}), then the  combination $|-\ra \equiv \frac{1}{\sqrt{2}}(|1\ra -|2\ra)$ decouples from the other orthogonal to it states. In the subspace of  states $|0\ra$ and 
\be
|+\ra \equiv\frac{1}{\sqrt{2}}(|1\ra +|2\ra),
\label{plus-def}
\ee
the Hamiltonian (\ref{do-3})  then reduces to a two state Hamiltonian 
\begin{equation}
  H_{LZ} = \left(
\begin{array}{cc}
\beta t & \sqrt{2}g \\
\sqrt{2} g & 0
\end{array}
  \right),
    \label{lz-eff1}
\end{equation}
which is the Hamiltonian of the standard two-state LZ model, which predicts the transition probability 
\be
P_{0\rightarrow +} =1-e^{-4\pi g^2/\beta}.
\label{popp}
\ee
Given the definition of $|+\rangle$, the transition probabilities to the original diabatic states are given by
\begin{equation}
P_{0\rightarrow 1}^{\varepsilon=0} =P_{0\rightarrow 2}^{\varepsilon=0}=\frac{1}{2} \left( 1-e^{-4\pi g^2/\beta}\right).
\label{P0-tr}
\end{equation}

Thus, we do find the symmetry at $\varepsilon=0$. To understand why the limit $\varepsilon \rightarrow 0$ is not equivalent to the strict equality $\varepsilon=0$, we note that the  probability of escaping from $|0\ra$ to {\it any} of the parallel levels behaves continuously. Namely, comparing (\ref{P0-tr}) with (\ref{DO3P}) we conclude that 
$$
P_{0\rightarrow 1}^{\varepsilon=0} +P_{0\rightarrow 2}^{\varepsilon=0}=P_{0\rightarrow 1} 
+P_{0\rightarrow 2},
$$
so, only the ratio (\ref{ratio1}) behaves discontinuously near $\varepsilon=0$. This suggests that the effective description by the two-state Hamiltonian (\ref{lz-eff1}) does capture a portion of physics. Indeed, the standard LZ model features a characteristic time-scale, 
$$
\tau_{LZ}=g/\beta,
$$
during which the nonadiabatic transitions from the original level saturate. This should be true even if we split the degeneracy, ensuring that  
 $\varepsilon \ll  1/\tau_{LZ}$. Such a small but nonzero value of $\varepsilon$ cannot accumulate its effect during $\tau_{LZ}$ and thus change the probability of escaping from the original level.  

\vspace{0.2cm}
\paragraph*{\it Effect of virtual transitions.}  In addition to the direct transitions  from $|0\ra$ to $|1\ra,\,|2\ra$, the system can produce virtual transitions between $|1\ra$ and $|2\ra$ via the state $|0\ra$ \cite{Yurovsky1999,Rangelov2005}. Such transitions do not change the population of $|0\ra$ but effectively couple $|1\ra$ and $|2\ra$. For $\beta t \gg g$, i.e., for $t\gg \tau_{LZ}$, this coupling can be estimated using  second order quantum perturbation theory. Namely, by assuming that $g/(\beta t) \ll 1$,  the diabatic states $|1\ra$ and $|2\ra$ can be redefined so that there are no nonzero  off-diagonal Hamiltonian matrix elements that  connect them to $|0\ra$. Up to the first order in $g/(\beta t)$, this redefinition is given by
\begin{eqnarray}
\nonumber |\psi_1 \rangle &=& |1\ra - \frac{g}{\beta t}|0\rangle +\frac{g^2}{2\varepsilon \beta t}|2\ra, \\
\label{new-states}
|\psi_2 \rangle &=& |2\ra - \frac{g}{\beta t}|0\rangle -\frac{g^2}{2\varepsilon \beta t} |1\ra.
\end{eqnarray}

Within the subspace~(\ref{new-states}), the matrix elements of the Hamiltonian (\ref{do-3}), are given by 
$$
V_{12}(t) =V_{21}(t) =\langle \psi_2|H |\psi_1\rangle =-\frac{g^2}{\beta t}, 
$$
where we used that $t>\tau_{LZ}$ and $\varepsilon \ll 1/\tau_{LZ}$.
The same virtual processes also renormalize the diabatic energies of $|\psi_1\ra$ and $|\psi_2\ra$ by 
\be
V_{11} =- \frac{g^2}{\beta t}, \quad V_{22} = -\frac{g^2}{\beta t}.
\label{en-diag1}
\ee
However, since the energies in (\ref{en-diag1}) are the same, we can gauge their effect away by  the phase shifts of the basis states as $|\psi_{1,2}\ra \rightarrow \exp\left(i\int^t g^2/(\beta \tau) \, d\tau \right) |\psi_{1,2}\ra$. Assuming that this is done, we find that for $t\gg \tau_{LZ}$ the effective Hamiltonian for interaction of the parallel diabatic levels via $|0\ra$ has the form
\begin{equation}
  H_{\rm eff} =\left( \begin{array}{cc}
-\varepsilon & -\gamma/t \\
-\gamma/t & \varepsilon 
\end{array}
  \right), \quad \gamma \equiv \frac{g^2}{\beta}.
    \label{Heff2}
\end{equation}
This effective Hamiltonian, acting in space of $|\psi_1\ra$ and $|\psi_2\ra$, should be supplemented by the initial condition that arises from the fact that after the direct transitions from the level $0$ saturate, that is, for $t\sim\tau_{LZ}$, the system emerges in the equal  superposition, $|+\ra$, which is also the eigenstate of the Hamiltonian (\ref{Heff2}) as $t\rightarrow 0_+$.

For early times, $t\sim \tau_{LZ}$, the initial state $|+\ra$ remains to be almost the eigenstate of $ H_{\rm eff}(t)$ due to the smallness of $\varepsilon$. Thus,
nonadiabatic transitions are negligible at that time. However, for
\begin{equation}
 t\sim \tau_{\varepsilon} \equiv \gamma/\varepsilon,   
 \label{tau-var}
\end{equation}
the off-diagonal coupling in $H_{\rm eff}$ is comparable to  $\varepsilon$. This leads to the nonadiabatic transitions between $|\psi_1\ra$ and $|\psi_2\ra$.
Note also that $|\psi_{1,2}\rangle$ asymptotically, as $t\rightarrow +\infty$, coincide with our original diabatic states $|1\ra,\, |2\ra$, so the final population probabilities in both bases coincide.

Thus, for very small $\varepsilon$, the evolution splits into two almost disjoint processes. The first process is essential in the interval $t\sim (-\tau_{LZ},\tau_{LZ})$, during which the probability of the transition  from $|0\ra$ to $|+\ra$ saturates, and this is described by the effective Hamiltonian (\ref{lz-eff1}). During much longer time $\sim \tau_{\varepsilon}$, defined in (\ref{tau-var}), the initially equal superposition $|+\ra$ dissolves, leading to the  asymmetric population distribution of $|1\ra$ and $|2\ra$, without changing the overall  probability to be in this subspace, and this process is described by the Hamiltonian 
(\ref{Heff2}). 

\vspace{0.2cm}
\paragraph*{\it Scattering due to the virtual transitions.} The Schr\"odinger equation (\ref{se}) with the Hamiltonian (\ref{Heff2}) is exactly solvable.  
Since $\tau_{\varepsilon}\gg \tau_{LZ}$, we can disregard the fact that the evolution with $H_{\rm eff}$ starts at finite time, and estimate the transition probability from $|+\ra$ to $|1\ra,\, |2\ra$ by assuming that the evolution with the Hamiltonian $H_{\rm eff}$ continues during time $t\in (0,+\infty)$. The expressions for the transition probabilities then have a particularly simple form (see e.g. \cite{Sinitsyn2014} for a simple derivation of the following formulas):
\be
P_{+\rightarrow 1} = \frac{1}{1+e^{-2\pi\gamma}}, \quad P_{+\rightarrow 2} = \frac{e^{-2\pi \gamma}}{1+e^{-2\pi \gamma}}, \quad \gamma \equiv \frac{g^2}{\beta}.
\label{pr-p12}
\ee

The independence of the probabilities in (\ref{pr-p12}) of $\varepsilon$ in this case is easy to prove. Specifically, for the Hamiltonian (\ref{Heff2}), if we rescale time in the Schr\"odinger equation (\ref{se}) as $t \rightarrow \lambda t$, where $\lambda >0$, we find the same equation but with 
different diabatic energies, such that $\varepsilon \rightarrow \lambda \varepsilon$. Since the integration interval $t \in (0,+\infty)$ is not affected by this time rescaling, the transition probabilities before and after the change of time variable should be the same. This implies that they must be independent of the parameter $\varepsilon$, as long as it remains positive.

We are now in a position to re-derive the transition probabilities (\ref{DO3P}) for the 3-state Demkov-Osherov model. 
The probability of turning from $|0\ra$ to $|1\ra,\, |2\ra$ is given by the product of the probability of transferring to the state $|+\ra$ during the Landau-Zener transition, which is given by Eq.~(\ref{popp}), times the probability of the state $|+\ra$ splitting into either $|1\ra$ or $|2\ra$, as described by Eq.~(\ref{pr-p12}):
\be
P_{0 \rightarrow 1}=P_{0\rightarrow +} \cdot P_{+\rightarrow 1}, \quad \quad P_{0 \rightarrow 1}=P_{0\rightarrow +} \cdot P_{+\rightarrow 2},
\ee
which reproduces Eq.~(\ref{DO3P}).

To summarize this section, the origin of the discontinuity of the transition probabilities in the limit $\varepsilon \rightarrow 0$ can be traced to the relevance of  a time-dependent energy scale, 
\be
E \sim \frac{g^2}{\beta t},
\label{new-scale}
\ee
which follows from the virtual transitions between the parallel levels via the state  $|0\ra$ having increasing  diabatic energy. The coupling $g$ and the time-dependence  of the diabatic energy of  $|0\ra$ can originate from forces that are much stronger than the $\varepsilon$-splitting. However, {\it altogether}, such forces  introduce a decaying  effective coupling (\ref{new-scale}), which at sufficiently large $t$ resonates with $\varepsilon$, and induces substantial nonadiabatic transitions between the nearly degenerate levels. 

For this mechanism to function as  described, the time of evolution with the Hamiltonian (\ref{do-3}) should terminate much later than 
$$
\tau_{\varepsilon}\equiv \frac{g^2}{\beta \varepsilon},
$$
so that the final energy of the state $|0\ra$ is inversely proportional to $\varepsilon$:
\be
E_{0}^{\rm fin} \gg \beta \tau_{\varepsilon} = \frac{g^2}{\varepsilon}.
\label{en-fin}
\ee

The physics of particle production in the driven Tavis-Cummings model is very similar to the dynamics of state amplitudes in the three-state example that we have just discussed. The main difference is that the elementary level crossing points will be replaced by critical points of phase transitions.

\section{Integrable MLZ models}
\label{integ-MLZ}
\subsection{Consistency conditions}
 The Schr\"odinger equation (\ref{se}) has a straightforward generalization to the evolution in multiple times: 
\begin{equation}
   i \frac{\partial}{\partial \tau_j} |\Psi( \bm{\tau}) \ra = H_j(\bm{\tau}) | \Psi ({\bm \tau}) \ra, \quad j=1, 2, \ldots, N, \quad N\ge 2,
\label{multi-time1}
\end{equation}
where $H_j$ are Hermitian operators and ${\bm \tau}\equiv (\tau_1,\ldots,\tau_N)$ is a vector of continuous real time variables.

The system in (\ref{multi-time1}) is overdetermined, which can be shown, e.g., by differentiating its $k$-th equation  by $\tau_j$ and the $j$-th equation by $\tau_k$. On the left hand side of these two equations we then find  the same expression; however, the right hand sides are the same only if the Hamiltonians satisfy  a {\it consistency condition}
\be
\left( \frac{\partial H_k}{\partial \tau_j} -\frac{\partial H_j}{\partial \tau_k}\right)-i [H_k,H_j] =0.
\label{cons-cond}
\ee
Only when Eq~(\ref{cons-cond}) is satisfied for all $k$ and $j$, does the system (\ref{multi-time1})  have a unique solution for all $\bm{\tau}$, except the Hamiltonian singularities.  The multi-time equation (\ref{multi-time1}) with conditions (\ref{cons-cond}) plays a key role  in the study of  integrable models  \cite{Faddeev1987,Fokas2006}. 

Recently, a novel application of this theory was discovered -- to identify and solve  MLZ models \cite{Sinitsyn2018}. For many MLZ Hamiltonians of the form (\ref{Hmlz}) one can identify another time variable and a different nontrivial Hamiltonian matrix such that the Schr\"odinger equation is extended to the system (\ref{multi-time1}). This can then be utilized to find the solution of the original Schr\"odinger equation (\ref{se}) analytically without approximations.

\vspace{0.2cm}
\paragraph*{\it Commuting $t/\tau$-pair.}
\label{ttau-sec}
All known fully solvable MLZ models are derived from the Hamiltonians that depend only linearly on the physical time $t$, and linearly or inversely linearly on a parameter $\tau$, which becomes an independent time variable when we extend the evolution to the two-time plane. A pair of such Hamiltonians that satisfies the integrability condition (\ref{cons-cond}) is called  a $t/\tau$-pair \cite{Chernyak2020}. 

For real symmetric $A$ in Eq.~(\ref{Hmlz}), the condition (\ref{cons-cond}) splits into two  simpler conditions: 
\begin{equation}
 \frac{\partial H_k}{\partial \tau_j} -\frac{\partial H_j}{\partial \tau_k}=0, \quad [H_k,H_j] =0.
    \label{cond2}
\end{equation}
As a second time variable, $\tau$, we choose some of the diagonal elements of the matrix $A$. For example, a parameter that rescales the parallel level splitting, $\varepsilon \rightarrow \tau \varepsilon$, in the Hamiltonian (\ref{do-3}). We can still treat $\tau$ as a constant for evolution along the true time but now we allow the time also to go along an arbitrary path in the two time plane $(t,\tau)$. The MLZ Hamiltonian then takes the form  
\begin{equation}
  H(t,\tau) =B t +B^{0} \tau+A_0,
    \label{Hmlz-2}
\end{equation}
where $B$ and $B^{0}$ are diagonal matrices, and $A_0$ does not depend on $t$ and $\tau$. The Schr\"odinger equation (\ref{se}) can then be extended to a system (\ref{multi-time1}), in which the second Hamiltonian is linear in $t$ but  generally has both linear and inverse linear dependence on $\tau$:
\begin{equation}
    H'(t,\tau) = B' \tau + B^{0} t +A' +\frac{C}{\tau},
    \label{h1-tt}
\end{equation}
with some diagonal $B'$, with the same $B^0$ as in (\ref{Hmlz-2}),  and with real symmetric $A'$ and $C$. 

For any given solvable MLZ model, the matrices $B,\,B^0,\, A$ can be read out from its Hamiltonian, and then the matrices $B',\, A', \, C$ are found by setting  the commutator, $[H,H']$, to zero and separately considering the equations linear in $t$ and $\tau$, as well as the equations without the time variables \cite{YUZBASHYAN2018323,Chernyak2020}.

\subsection{Integrability of  3-state Demkov-Osherov model}
For example, let us expand the model~(\ref{do-3}) to the two-time evolution. We first introduce a new variable $\tau$ that rescales the splitting between the parallel levels: 
\begin{equation}
   H(t,\tau) =\left(
   \begin{array}{ccc}
    \beta t & g &g \\
    g & -\tau\varepsilon &0  \\
    g & 0  & \tau \varepsilon
   \end{array}
   \right), \quad \beta,\, \varepsilon>0.
    \label{do-31}
\end{equation}
We then find the commuting matrix of the form (\ref{Hmlz-2}) for it, which is
\begin{equation}
   H'(t,\tau) =\left(
   \begin{array}{ccc}
    \frac{\varepsilon^2}{\beta} \tau -\frac{g^2}{\beta \tau} & -\frac{\varepsilon g}{\beta} &\frac{\varepsilon g}{\beta} \\
    -\frac{\varepsilon g}{\beta} &\varepsilon t &-\frac{g^2}{\beta \tau}  \\
    \frac{\varepsilon g}{\beta} & -\frac{g^2}{\beta \tau}  & - \varepsilon \tau
   \end{array}
   \right), \quad \beta,\, \varepsilon>0.
    \label{do-32}
\end{equation}

  \begin{figure}[t!]
    \centering
    \includegraphics[width=0.45\textwidth]{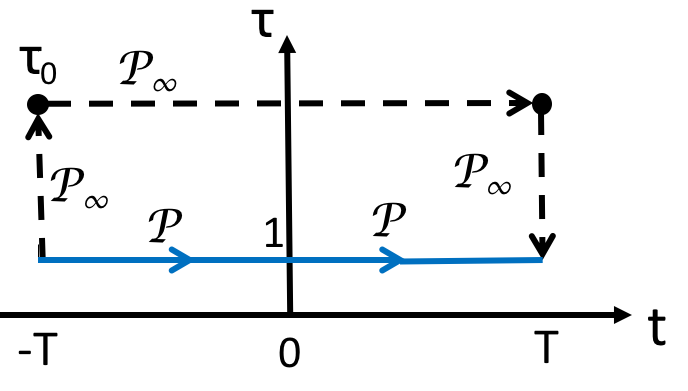}
    \caption{An integration path ${\cal P}$ (blue arrows) with $\tau=1$ and $t\in (-T,T)$, where $T\rightarrow \infty$, is deformed into a path ${\cal P}_{\infty}$, such that the horizontal part of ${\cal P}_{\infty}$ has $\tau=\tau_0\ne 1$ (dashed black arrows). This path deformation was proposed in \cite{Sinitsyn2018,Malla2022}. The path deformations do not change the evolution matrix, as the initial and final path points remain the same. The vertical legs of ${\cal P}_{\infty}$ have $t=\pm T $, so they contribute only to the trivial adiabatic phases in the evolution matrix, and do not affect the transition probabilities.}
    \label{paths-fig}%
\end{figure}
 
\vspace{0.2cm}
\paragraph*{\it  Time-path deformations.} Let us define the evolution operator
 \be
 U=\hat{\cal T}_{\cal P}\exp \left( -i \int_{\cal P} H(t,\tau)\, dt+H'(t,\tau) \, d\tau \right),
 \label{path1}
 \ee
 where $\hat{\cal T}_{\cal P}$ is the path ordering operator along  a path ${\cal P}$ that connects the points $(-\infty,1)$ and $(+\infty,1)$ in the two-time space $(t,\tau)$.
  The integrability conditions imply that the non-Abelian gauge field with components ${\bf A}(t,\tau) = (-H,-H')$ has zero curvature. This, in turn, means  that the result of integration in (\ref{path1}) does not change after smooth deformations of ${\cal P}$  that keep only the initial and final points of ${\cal P}$ intact \cite{Sinitsyn2018,Chernyak2020}, while avoiding singularities of the $\tau$-dependent Hamiltonians, as we show in Fig.~\ref{paths-fig}.

For our 3-state Demkov-Osherov model, the physical evolution corresponds to the changes of $t$ from $-\infty$ to $+\infty$ at $\tau=1$. Then, ${\cal P}$ starts at the point $(t,\tau)=(-\infty,1)$.  We are free to fix, initially, $t$ and change $\tau$ from this point to another value, $\tau_0$, and only then perform the $t$-evolution at fixed $\tau=\tau_0$. After this, we  can return $\tau$ to $\tau=1$ at $t=+\infty$, as shown in Fig.~\ref{paths-fig}. 
 
 The $\tau$-evolution at fixed $t=- \infty$ or $t=+\infty$ is purely adiabatic due to the quadratic dependence of the diagonal elements of $H'$ on $t$. Therefore, the transition probability matrices of the 3-state LZ-models that differ  only by  $\tau>0$ are identical.

In all known fully solvable MLZ models, the commuting pair of the Hamiltonians has the form (\ref{Hmlz-2}) and (\ref{h1-tt}), in which the matrix $B_0$ commutes with $B$, and therefore is diagonal in the diabatic basis. Hence, $B^0$ in (\ref{Hmlz-2}) represents the $t$-independent components of the diabatic energies. We have just shown that,  for such MLZ models, the transition probabilities do not depend on the rescaling parameter $\tau$, which controls relative values for such couplings. For example, for the 3-state Demkov-Osherov model, this means that the transition probabilities do not depend on the parallel level splitting parameter $\varepsilon$.

\vspace{0.2cm}
\paragraph*{\it Independent crossing ``approximation".} Knowing such facts, the transition probabilities are easy to derive for any solvable MLZ-model. Thus, for the Hamiltonian (\ref{do-3}), we can find the transition probabilities by assuming that   $\varepsilon \gg g,\, \beta^{1/2}$. Then, the two avoided crossing points, at the crossings of level $0$ with levels $1$ and $2$, are well-separated in time-energy space. Hence, one can apply the LZ formula  in order to obtain the microstate probabilities after each pairwise diabatic level crossing. Sometimes, this procedure is used as an approximation when all avoided crossings are well-separated. However,  for the solvable MLZ models its predictions become analytically exact.

For instance, consider that in Eq.~(\ref{do-3}) the parameter $\varepsilon$ is large. In Fig.~\ref{DO-fig}, the parallel levels would be then at a large distance from each other.  By disregarding the state $|2\ra$, the effective Hamiltonian in space of states $|0\ra$ and $|1\ra$ near the first crossing point in Fig.~\ref{DO-fig} is a standard two-state LZ system: 
\begin{equation}
H_1=\left( \begin{array}{cc}
\beta t &g \\
g& -\varepsilon
\end{array}\right).
\label{lz-ham}
\end{equation}
The probability of staying at level $0$ after crossing the first resonance is given by the LZ formula:
\begin{equation}
P_{0\rightarrow 0}^{(1)}=e^{-2\pi g^2/\beta}.
\label{LZ}
\end{equation}
The probability of turning at this point to level $1$ is given by $P_{0\rightarrow 1}^{(1)}=1-e^{-2\pi g^2/\beta}$, and this coincides with the corresponding expression in Eq.~(\ref{DO3P}) because level $1$ does not encounter resonances in the future. 
However, level $0$ later experiences  the crossing with level $2$.  This crossing, for large $\varepsilon$, is also described by the standard LZ formula, with the probability of staying at level $0$ again gaining a factor, $P_{0\rightarrow 0}^{(2)}=e^{-2\pi g^2/\beta}$. The entire probability of staying at level $0$ after passing through both resonances is given by
$$
P_{0\rightarrow 0}=P_{0\rightarrow 0}^{(1)} \cdot P_{0\rightarrow 0}^{(2)}=e^{-4\pi g^2/\beta},
$$
which coincides with the analogous expression in (\ref{DO3P}). Finally, the probability of ending up at level $2$ is the probability of staying at level $1$ at the first resonance and then turning to level $2$:
$$
P_{0\rightarrow 1}=P_{0\rightarrow 0}^{(1)} \cdot P_{0\rightarrow 2}^{(2)}=e^{-2\pi g^2/\beta}\left(1-e^{-2\pi g^2/\beta}\right),
$$
which coincides with the analogous expression in (\ref{DO3P}).

{\it Landau-Zener formula from MLZ integrability.} Finally, we mention a curious recent observation, made by C. Sun \cite{Sun2025}, that the LZ formula for two levels (\ref{LZ}) is a consequence of the MLZ integrability. It is easy to illustrate with  our 3-state Demkov-Osherov model example. Indeed, we have just proven that, the survival probability $P_{0\rightarrow 0}$ does not depend on $\varepsilon$. In addition, the survival probability for the LZ Hamiltonian (\ref{lz-ham}) must depend only on the ratio 
$$
\gamma\equiv g^2/\beta,
$$
because, by rescaling time $t\rightarrow t/\sqrt{\beta}$, we leave only this parameter combination in the time-dependent Schr\"odinger equation. Then, on one hand, we have already discussed that, for $\varepsilon \gg g$, the initial state $|0\rangle$ passes through two distant elementary LZ resonances, so we must have 
$P_{0\rightarrow 0}=\left(P_{0\rightarrow 0}^{(1)}(\gamma) \right)^2$.
On the other hand, at $\varepsilon =0$, $P_{0\rightarrow 0}$ is the 
survival probability in the LZ Hamiltonian (\ref{lz-eff1}) that differs from Eq.~(\ref{lz-ham}) merely by replacing $g$ for $g\sqrt{2}$. By equating $P_{0\rightarrow 0}$ obtained at $\varepsilon =0$ and for $\varepsilon \gg g$, we find that 
$$
P_{0\rightarrow 0}^{(1)}(2\gamma)=\left(P_{0\rightarrow 0}^{(1)}(\gamma)\right)^2,
$$
which has a solution $P_{0\rightarrow 0}^{(1)}=e^{c\gamma}$, where the constant $c=-2\pi$ can be calculated perturbatively and trivially in the limit of large $ \beta$. Knowing the solution of the Demkov-Osherov model, the transition probabilities~(\ref{pr-p12}) for the model with decaying coupling can be derived from the Demkov-Osherov model in the limit $\varepsilon \rightarrow 0$, as we have discussed; therefore, they can be considered as the consequence of the MLZ integrability too. Analogously, solvable many-body models, with decaying as $\propto 1/t$ couplings, can be obtained from the driven Tavis-Cummings model in a similar limit. We will show a specific example in Section~\ref{PT}.

\subsection{Integrable bosonic driven Tavis-Cummings model}

We have worked out the 3-state Demkov-Osherov model in detail because the same features and calculation tricks are used to solve much more complex models, which sometimes have  combinatorially large phase spaces and describe intricate many-body processes. Now, we consider the most complex MLZ model that has been solved to date. 

\vspace{0.2cm}
\paragraph*{\it Integrability conditions for interacting bosons.}
Let  $\hat{a}_k$ and $\hat{b}_k$, where $k=1,2,\ldots$, as well as $\hat{\Psi}$ be the standard commuting bosonic annihilation operators.
We construct  operators
$$
K_k^+\equiv \hat{a}_k^{\dagger}\hat{b}_k^{\dagger}, \quad K_k^-\equiv \hat{a}_k\hat{b}_k, \quad \hat{q}_k\equiv \hat{a}^{\dagger}_k\hat{a}_k+\hat{b}^{\dagger}_k\hat{b}_k,
$$ which satisfy 
\be
[\hat{q}_k,K^{\pm}_k]=\pm 2K_k^{\pm}, \quad [K_k^-,K_k^+]=(\hat{q}_k+1).
\label{com1}
\ee
The driven bosonic Tavis-Cummings Hamiltonian is defined by
\begin{eqnarray}
\label{h23}
\nonumber H(t,\tau)&=&\beta t \hat{\Psi}^{\dagger} \hat{\Psi} \\
&+&\sum_k  \left\{ \tau \varepsilon_k   \hat{q}_k/2+g\left[\hat{\Psi}^{\dagger}K^-_k  +\hat{\Psi}K^+_k \right] \right\},
\end{eqnarray}
where the sum over $k$ is over an arbitrarily large number of ``reaction channels" that describe splittings of particles $\Psi$ into the pairs of bosons, $a_k$ and $b_k$.
We assume that the time-evolution during $t\in (-\infty,\infty)$ at constant $\tau=1$ corresponds to the physical situation that we want to ultimately understand. 

This time-dependent model is integrable because $H(t,\tau)$ in Eq.~(\ref{h23}) commutes with 
\begin{eqnarray}
\label{h23-c}
\nonumber H'(t,\tau)&=&\sum_k \big \{ \varepsilon_k (t-\frac{\tau \varepsilon_k}{\beta})\frac{\hat{q}_k}{2}   -\frac{g \varepsilon_k}{\beta}\left[\hat{\Psi}^{\dagger}K^-_k  +\hat{\Psi}K^+_k \right] \big \} 
\\
&+&\frac{g^2}{\beta\tau}\sum_{i,j;\, i\ne j} \left(K_i^+K_j^- -(\hat{q}_i+1)(\hat{q}_j+1)/4\right),
\end{eqnarray}
and these  two Hermitian operators satisfy the additional condition
$$
\partial H/\partial \tau = \partial H'/\partial t.
$$
We will consider the initial situation, as $t\rightarrow -\infty$, with no bosons in the modes $\hat{a}_k$ and $\hat{b}_k$, but arbitrary $N$ bosons in $\hat{\Psi}$.

\vspace{0.2cm}
\paragraph*{\it Transition probabilities in the multi-channel reaction.} Consider now the multi-channel case, $k=1,2,\ldots,s$. As the operators in (\ref{h23}) and (\ref{h23-c}) are a $t/\tau$-pair of the form (\ref{Hmlz-2})-(\ref{h1-tt}), the integrability allows the deformation of the time-integration path as in Fig.~\ref{paths-fig}, so that we initially change $\tau$ from $\tau=1$ to a very large value at $t=-T\rightarrow -\infty$. The Hamiltonian for this piece of the evolution is $H'$ in Eq.~(\ref{h23-c}). Due to the large value of $|t|$, this piece of evolution  is adiabatic and does not change the probabilities of the microstates. We then change $t$ during $(-T, +T)$ at the fixed new very large value of $\tau$. This time evolution proceeds with the Hamiltonian $H(t,\tau)$ that has all resonance energies $\varepsilon_k$ rescaled by $\tau$ from their physical values. Finally, we move towards the point at $(t,\tau)=(+\infty,1)$. This final piece of the evolution is again adiabatic due to the large value of $|t|$. Since the pieces of the evolution over $\tau$ did not change the microstate probabilities, we find that the time evolution with $t\in (-\infty,+\infty)$  for different $\tau>0$ leads to the same transition probabilities.  

Knowing about the independence of the transition probabilities of  $\tau$, we set $\tau \gg 1$, so that the energy distance between the reaction resonances is much larger than the scattering amplitude $\sim g \sqrt{N}$. We then treat each resonance independently in the chronological order.
As in the Demkov-Osherov model, we then find that the transition probabilities, from the initial molecular ground state to an arbitrary final microstate,
do not depend on the values of the chemical potentials $\varepsilon_k$, except their ordering in magnitude. In what follows, we will choose indices so that $\varepsilon_1<\varepsilon_2<,\ldots$.

As in the Demkov-Osherov model, the integrability allows us to find the transition probabilities recursively. Namely, suppose the system has $s+1$ reaction channels, so that the final microstate is characterized by $n_1,\ldots, n_{s+1}$ atomic pairs in the corresponding energy surfaces.
The probability of this state is the same as to generate $n_1, \ldots n_{s}$ pairs of atoms in the first $s$ channels multiplied by the probability of generating $n_{s+1}$ pairs starting from $N-\sum_{k=1}^{s}n_k$ of $\Psi$-bosons entering the $(s+1)$-st resonance. We refer to \cite{Malla2022} for further details of the derivation of the following recursive formula that connects the probabilities in models with $(s+1)$ and $s$ reaction channels. Let
\be
x \equiv e^{-2\pi g^2/\beta},
\label{x-def}
\ee
and  
\be
\nonumber (a,x)_q\equiv \prod_{k=0}^{q-1} (1-a^kx)=(1-a)(1-ax)\ldots (1-a^{q-1}x),
\label{QP-def}
\ee
which is a special function called q-deformed Pochhammer symbol. Then,
\begin{equation}
P_{n_1,\ldots, n_s,n_{s+1}}\!=\!P_{n_1,\ldots, n_s} x^{N-\sum \limits_{k=1}^s n_k}\left(x^{N+1-\sum_{k=1}^{s+1} n_k },x\right)_{n_{s+1}}.
\label{P-itter}
\end{equation}
By setting $P=1$ for the case with no reaction channels, $s=0$, this formula can be used to  inductively reconstruct the formulas for the transition probability to arbitrary microstate.

\vspace{0.2cm}
\paragraph*{\it Coherent path to thermalization.} For  the complete molecular dissociation into the atomic pairs, which is found for infinitely many reaction channels ($s=\infty$), the final  probability distribution of the atoms over the energy of different final atomic states was found to be the perfect Gibbs distribution for the non-interacting bosons. Namely, consider the equidistant spectrum of the final atomic states: $\varepsilon_k  =\varepsilon k$, where $k=1,2,\ldots, \infty$, are the indices of the reaction channels, as in Eq.~(\ref{P-itter}). Then, the probability of finding  $n_1,n_2\ldots$ pairs of the bosonic atoms in these atomic modes with different energies is \cite{Malla2022}
\be
P_{n_1,n_2,\ldots} \propto e^{-\frac{1}{k_BT} \sum_{k} \varepsilon_k n_k} \delta \left(N-\sum_k n_k \right),
\label{pgen}
\ee
where the  temperature is proportional to the sweep rate:
\begin{equation}
k_BT =\frac{\beta \varepsilon}{2\pi g^2 }. 
\label{kt1}
\end{equation}

\section{Single channel reaction}
\label{single-ch}
\subsection{Transition probabilities}
The general solution of the model (\ref{h23}) can be derived from the solution for only a single channel reaction. Consider  the case with index $k$ in the sum having only one value, $1$. The Hamiltonian  can then be simplified to
 \begin{equation}
   \label{H2-mol-2}
 H(t)=\beta t \hat{\Psi}^{\dagger} \hat{\Psi}  +g\left(\hat{\Psi}^{\dagger}\hat{a}\hat{b}  +\hat{\Psi}\hat{a}^{\dagger}\hat{b}^{\dagger} \right).  
 \end{equation}
Here, $\hat{a}$ and $\hat{b}$ are the bosonic annihilation operators of the two atomic modes, and $\hat{\Psi}$ is the bosonic operator for annihilation of the ``molecules", i.e., the particles that can split into two atoms.  The coupling $g$ characterizes the reaction rate at the Feshbach resonance. The initial conditions correspond to $N$ molecules and no  atoms  as $t\rightarrow -\infty$.

The probability of splitting $m$, out of $N$, molecules to the pairs of atoms $a$ and $b$ can be obtained from Eq.~(\ref{P-itter}) by identifying $m$ with $n_1$ and assuming that $n_k=0$ for $k\ge 2$:
\be
P_{m}=x^{N-m}(x^{N-m+1},x)_{m},
\label{P-one}
\ee
where $x\equiv e^{-2\pi g^2/\beta}$ is the same as defined in Eq.~(\ref{x-def}).
Equation~(\ref{P-one}) was derived in \cite{Sun2016,Sun2019}, for a different physical interpretation of the model.

\subsection{Quasi-adiabatic regime}
By {\it quasi-adiabatic} we mean here that the transition through the resonance is sufficiently slow to dissociate a considerable fraction of the molecules but the probability of ending up in the new ground state may still be negligible. This corresponds, in our case, to the sweep rates such that 
$$
\frac{\ln N}{N} \ll \frac{2\pi g^2}{\beta} \ll 1.
$$
In our model, this regime corresponds to negligible probability of the event with $m=0$, i.e., $P_0<1/N$. 
In this regime, 
it is also convenient to work not with the numbers $m$, but rather with the deviation 
$$
\nu=N-m.
$$

Then, using  properties of the Q-Pochhamer symbol, we  rewrite
$$
P({\nu})\equiv P_{N-\nu}=x^{\nu} (x^{\nu+1},x)_{N-\nu}=x^{\nu} \frac{(x,x)_N}{(x,x)_{\nu}}.
$$
In the quasi-adiabatic regime, we can safely approximate $(x,x)_N$ by $(x,x)_{\infty}$, which leads to the approximation of $P(\nu)$ by the Euler distribution \cite{Sun2016}: 
\be
P({\nu})\approx x^{\nu} \frac{(x,x)_{\infty}}{(x,x)_{\nu}}.
\label{p1-eu}
\ee

More generally, the Euler distribution depends on one additional parameter, $\alpha$, so that up to the normalization factor 
\begin{equation}
P_E(\nu)\sim \alpha^{\nu}/(x,x)_{\nu}.
\label{p-euler}
\end{equation}
Equation~(\ref{p1-eu}) corresponds to Eq.~(\ref{p-euler}) at $\alpha=x$.

The average of this distribution is known \cite{Sun2016}:
\be
\la \nu \ra \equiv \sum_{\nu=0}^{\infty}\nu P_E(\nu) =\sum_{j=0}^{\infty} \frac{\alpha x^j}{1-\alpha x^j}.
\label{eu-av}
\ee
This sum is convenient to express
via the {\it q-digamma function}
\be
\psi_q(z)=-\ln(1-q)+
\ln q\sum_{n=0}^{\infty}
\frac{q^{n+z}}{1-q^{n+z}}.
\label{digamma-def}
\ee

In the case of Eq.~(\ref{p1-eu}), $\alpha=x$, so
\be
\la \nu \ra=\frac{\ln(1-x)+\psi_x(1)}{\ln x}.
\label{nu1-av}
\ee

We now recall that from the definition $x=e^{-2\pi g^2/\beta}$ it follows that $0<x<1$, and  $x\approx 1$ in the considered quasi-adiabatic regime. The value of the $q$-digamma function as $q\rightarrow 1$ is known: 
$$
\psi_1(1)=-\gamma_E=-0.57721\ldots,
$$
where $\gamma_E$ is known as Euler's constant. In terms of the original parameters we then find
\begin{eqnarray}
\label{nu1-av-fin}
\nonumber \la \nu \ra&=&\frac{
\beta}{2\pi g^2}
\left[ \gamma_E -\ln\left(1-e^{-2\pi g^2/\beta }\right)\right] \\
&\approx&
\frac{\beta}{2\pi g^2}
\left[ \ln\left(\frac{\beta}{2\pi g^2}\right)+\gamma_E \right].
\end{eqnarray}

In the quasi-adiabatic regime, the ratio $\beta/(2\pi g^2)$ can be as large as $\sim N/\ln N$, hence the logarithmic contribution in (\ref{nu1-av-fin}) is considerably larger than $\gamma_E$.

The almost-linear scaling of $\la \nu \ra$  with $\beta$ in (\ref{nu1-av-fin}) reflects the fact that the system during its evolution passes through a second-order quantum phase transition. The molecular dissociation increases the number of bosons in the system, which leads to a positive feedback that forces more molecules to dissociate. On one hand, this  explains why far from the true adiabatic limit ($g^2/\beta \ll 1$) the number of non-dissociated molecules, $\nu$, is small, i.e. the reaction behaves almost as an adiabatic process.  On the other hand, this regime is distinct from the adiabatic regime because the number of the nonadiabatic excitations  is suppressed by small $\beta$ non-exponentially but rather as the power law $\la \nu \ra \sim \beta \ln \beta$.

\subsection{Critical sweep rate}

\begin{figure*}[t!]%
\centering \includegraphics[width=1.8\columnwidth]{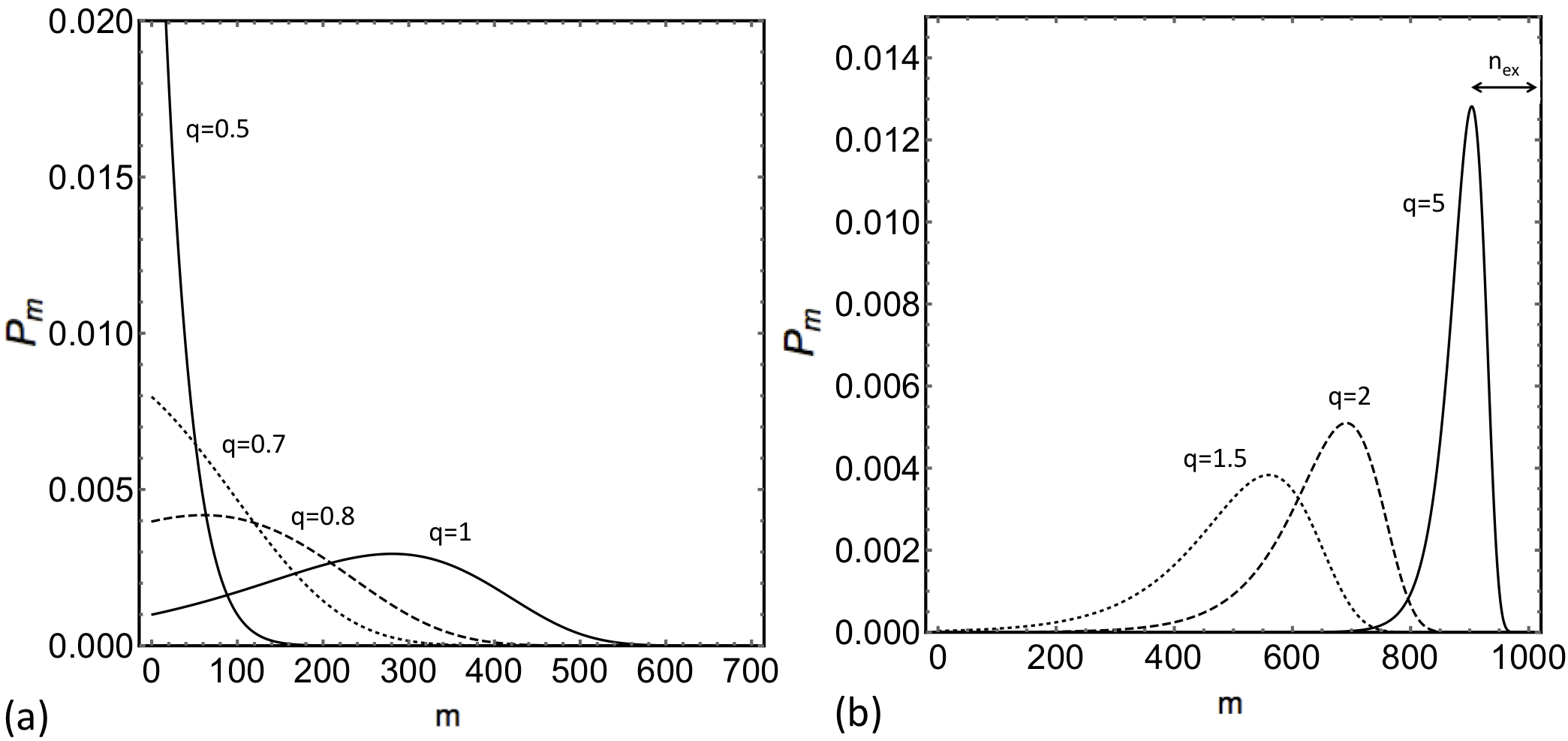}
\caption{The distribution of the number of the dissociated molecules, $m$, according to Eq.~(\ref{P-one}) at the initial number of molecules $N=1000$. Different curves correspond to different values of the parameter  $
q\equiv \frac{2\pi g^2N}{\beta \ln  N}
$. The slower sweep rates $\beta$ correspond to the distributions with larger $q$.
For different $N \gg 1$ but the same $q$, the distributions would have almost the same shapes as for the chosen here value, $N=1000$.  
(a) For $q<q_c\approx 0.8$ the distribution has its maximum at $m=0$, which means that the molecules resist the dissociation. However, for $q>q_c$, the distribution maximum shifts to a finite $m\ne 0$. (b) For $q\gg 1$, the maximum is very close to the boundary value of the complete dissociation at $m=N$, although $P_N$ remains very small up to $q\sim N$. Already for $q>5$, the distribution  is  sharply peaked, and the distance of its maximum to the boundary at $m=N$ defines the characteristic  number of the nonadiabatic excitations, $n_{ex}$. The asymmetry of the excitation distribution for $q>1$ is responsible for a noticeable difference between the position of the maximum, Eq.~(\ref{nex}), and the average number of excitations, Eq.~(\ref{nu1-av-fin}). }
\label{Pn-fig}
\end{figure*}

Reference~\cite{Sun2019} has shown that the entire  distribution (\ref{P-one}) in Fig.~\ref{Pn-fig}, for varying values of the combination
$$
q\equiv \frac{2\pi g^2N}{\beta \ln  N},
$$
experiences a phase transition. 
The truly adiabatic regime corresponds to $P_N \approx 1$, which requires $x\sim 1$ (as defined in Eq.~(\ref{x-def}), where the sweep rate is $\beta \sim 2\pi g^2$. Now, consider what can occur at a much faster sweep rate, on the order of $\beta \sim g^2N/ \ln N$.

\vspace{0.2cm}
\paragraph*{\it The maximum of the probability distribution.}
For fast sweeps, the distribution (\ref{P-one}) is peaked at $m=0$ (see Fig.~\ref{Pn-fig}(a)). We can say then  that this is the transition regime in which the molecules resist the dissociation. At some critical rate, however, the maximum of the distribution (\ref{P-one}) appears at $m>0$. In the thermodynamic limit, $N\rightarrow \infty$, this is the regime with a finite fraction of molecules splitting into the atoms. 

We can define the critical $\beta$ that separates the two regimes of the process by requiring that it is $\beta=\beta_c$ for which the  maximum of the distribution (\ref{P-one}) moves from $m=0$ to $m=1$, i.e., $P_0=P_1$. Using (\ref{P-one}), this condition  leads to a condition on the critical value of  $x=x_c \equiv e^{-2\pi g^2/\beta_c}$:
\begin{equation}
x_c=(1-x_c^N).
\label{cond-mol1}
\end{equation}
We are interested in the case where $N \gg 1$. Hence, the condition (\ref{cond-mol1}) must be satisfied for $\alpha \equiv 2\pi g^2/\beta_c  \ll 1$. We can then approximate $x_c=e^{-\alpha} \approx 1-\alpha$, which leads us to 
$$
\alpha = e^{-\alpha N}. 
$$
Let us search for the solution of this equation in the form $\alpha =\frac{q \ln  N}{N}$. We find $q=\frac{N}{\ln  N} e^{-q\ln  N}$. Note that if we set $q=1$, then we do not satisfy this equation but we cancel the leading order in $N$ term in it. Hence, let us look for the solution in the form $q=1-\delta$, where $\delta \ll 1$. We  then find
\begin{equation}
1-\delta = \frac{e^{\delta \ln  N}}{\ln  N}. 
\label{interm1}
\end{equation}
Again, treating $N$ as very large, we can disregard $\delta $ on the left hand side and then find 
\begin{equation}
\delta \approx \ln \ln  N/\ln  N, 
\label{firstapprox}
\end{equation}
which  decays to zero as $N\rightarrow \infty$ but extremely slowly, so it is better to keep  one additional correction to $q$ in the formulas. It is found by substituting (\ref{firstapprox}) into the left hand side of (\ref{interm1}).  
 The critical combination of the parameters then corresponds to
$$
\alpha = \frac{\ln  N}{ N} \left(1-\frac{\ln (\ln  N-\ln \ln  N)}{\ln  N} \right).
$$
In experiments, the readily controllable parameter is the sweep rate $\beta$. From $\alpha \equiv 2\pi g^2/\beta_c$, the critical value of the sweep rate for $N\gg 1$ is  given by 
\begin{equation}
\beta_c\approx\frac{2\pi g^2N}{\ln N-\ln (\ln  N-\ln \ln  N) }.
\label{betac1}
\end{equation}
The corrections to this formula decay  with growing $N$ very slowly. For experimentally relevant values $N\sim 10^2-10^6$, this formula reproduces the numerically exact value of $\beta_c$  only up to one-two significant digits. This is sufficient, however, to deduce that the dynamic phase transition occurs at the sweep rates considerably larger, by a factor $\sim N/\ln  N$, than the adiabatic value $\beta \sim 2\pi g^2$.

\vspace{0.2cm}
\paragraph*{\it The nonadiabatic excitations.} Let us now consider the condition that the peak of the distribution (\ref{P-one}) is in between $m$ and $m+1$:
$$
P_{m}=P_{m+1}.
$$
Substituting the definition of the q-Pochhammer symbol  here, many terms again cancel, leading to 
\begin{equation}
    x=1-x^{N-m}.
    \label{xnex}
\end{equation}
We remind that $m$ here is the number of the molecules that split into the atoms. Therefore, $N-m$ is the number of molecules that survive the transition through the Feshbach resonance. If the evolution were adiabatic, all molecules would dissociate, so we can interpret 
$$
n_{ex} \equiv N-m
$$
as the number of the nonadiabatic excitations.  We can associate the peak of the probability distribution of $n_{ex}$ with 
the characteristic number of the excitations. This definition, along with $x=e^{-2\pi g^2/\beta}$, used with Eq.~(\ref{xnex}), leads us to the expression for the effective number of the nonadiabatic  excitations \cite{Sun2019},
\begin{equation}
n_{ex} =-\frac{\beta \ln  (1-e^{-2\pi g^2/\beta})}{2\pi g^2}.
\label{nex}
\end{equation}
Up to the subdominant contribution $\propto\gamma_E$ in (\ref{nu1-av-fin}), this expression coincides with $\la \nu \ra$ for $\nu=N-m$. This is not surprising as the position of the peak of the distribution almost coincides with the average of this distribution.

For  $\beta$ larger but of the order of the critical value $\beta_c$, and for large $N$, we have $2\pi g^2/\beta \ll 1$. Hence, we arrive at the scaling of the nonadiabatic excitations for slow sweeping rates $\beta$ \cite{Itin2009,Sadhasivam2024}:
\begin{equation}
    n_{ex} \approx \Gamma \ln  (\Gamma), \quad \Gamma \equiv \beta/(2\pi g^2).
    \label{x-nex-scal}
\end{equation}

We interpret this as the power law, $n_{ex} \propto \beta$, up to the logarithmic correction. This law persists for $\beta$ smaller than $\beta_c \sim 2\pi g^2 N/\ln  N$ and terminates at the slowest rate $\beta \sim 2\pi g^2$, which corresponds to the onset of the truly adiabatic regime with exponentially suppressed excitations.

\vspace{0.2cm}
\paragraph*{\it Dynamic phase transition.} 
The expression for the position of the  distribution maximum, $m_{\rm max} \equiv N-n_{ex}$ is given by:
\be
m_{\rm max} =
\begin{cases}
N+\frac{\beta \ln  (1-e^{-2\pi g^2/\beta})}{2\pi g^2},& \text{if } \beta \leq \beta_c,\\
    0,              & \text{otherwise}.
\end{cases}
\label{peak}
\ee
It is almost an exact result, limited only by the assumption of large $N$. Hence, it is valid not only in the quasi-adiabatic regime but generally for $m_{\rm max}>0$, i.e., for all $\beta<\beta_c$. Therefore, we can investigate the behavior of $m_{\rm max}$ near the critical sweep rate at which $m_{\rm max}\ll N$. The critical rate corresponds to 
$m_{\rm max}(\beta_c)=0$, near which the slowly changing expression in the logarithm in (\ref{peak}) can be approximated by 
$$
\ln \left(1-e^{-2\pi g^2/\beta}\right) \approx -\frac{2\pi g^2 N}{\beta_c}. 
$$
Substituting this back to (\ref{peak}), we find that near the critical sweep rate,  the position of the peak behaves as 
\begin{equation}
m_{\rm max} =N\left(1-\frac{\beta}{\beta_c} \right)^{\mu}, \quad \mu=1, \quad \beta<\beta_c,    
\end{equation}
where $\mu$ is a characteristic exponent, whose value we extracted from the exact solution. 

Thus, the exact solution of the model~(\ref{H2-mol-2}) reveals a dynamic critical phenomenon. The
critical point here refers to a strongly nonadiabatic regime, in which the role of the control parameter is played by the sweep rate $\beta$. 
In this unusual phase transition, the system is explicitly time-dependent. Nevertheless, in the thermodynamic limit, $N\rightarrow \infty$, certain characteristics of the emerging state after the sweep can depend  as a power law on $\beta$ near its critical value.

\subsection{Scattering phases}
\begin{figure}[t!]
\centering \includegraphics[width=0.65\columnwidth]{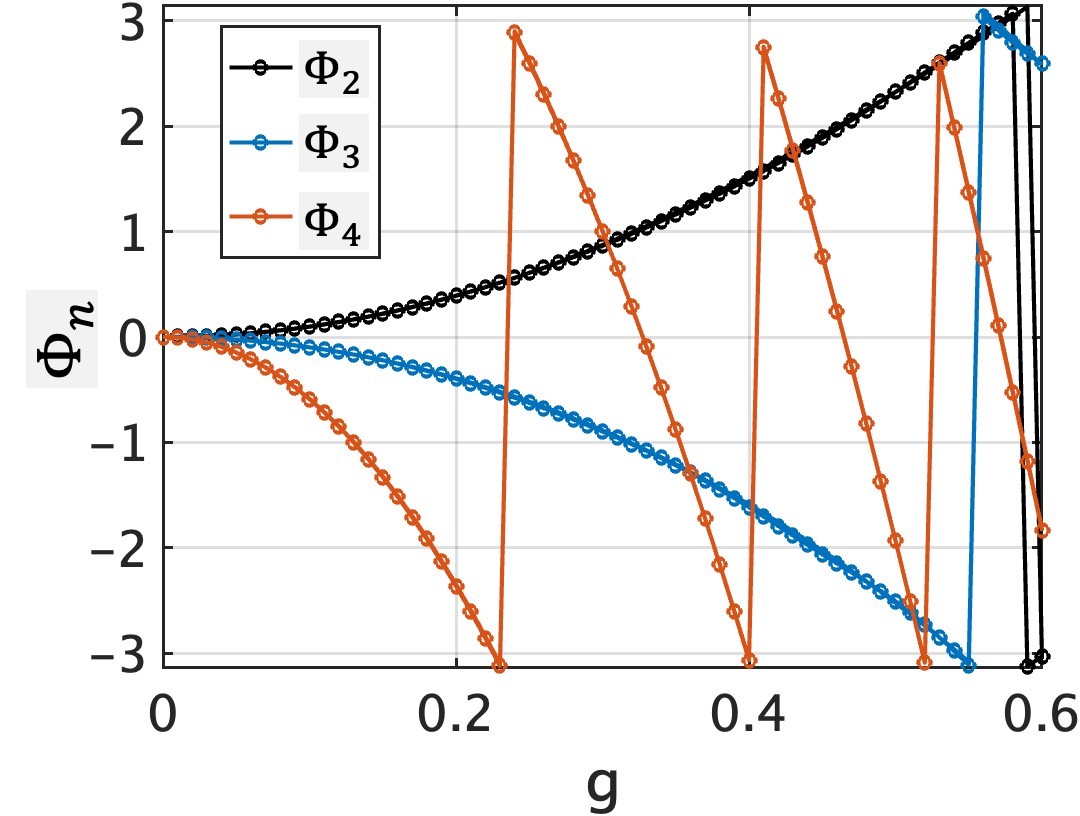}
\caption{Numerical test for Eq.~(\ref{phase-degenerate}) in 4-state Tavis-Cummings model ($N=3$). Phases were obtained for evolution over time $t\in (-100,100)$ at $\beta=1$. Here, we normalized $\Phi_n\equiv \Phi^{n}(g)-\Phi^{n}(g\rightarrow 0)$ and plotted these phases as functions of the coupling parameter $g$ for $n=2$ (black), $n=3$ (blue), and $n=4$ (red). The solid curves represent the analytical predictions, while the discrete points denote the results of the numerical solution of the time-dependent Schr\"odinger equation.
}
\label{phase3degenerate}
\end{figure}

The single-channel chemical reaction also creates nontrivial quantum correlations, which are fully quantified using scattering phases as well as the transition probabilities associated with the final wavefunction. The eigenstates of Tavis-Cummings Hamiltonian (\ref{H2-mol-2})
behave on time asymptotically as 
\begin{equation}
|\psi^m_{\pm}(t)\rangle = e^{i\phi^m_{\pm}(t)}|m\rangle, \quad \text{as} \quad t \to \pm \infty, \quad m = 0, \ldots, N,
\end{equation}
where the phases are given by
\begin{equation}
\phi^m_{\pm}(t) = -\frac{m\beta t^2}{2} +\frac{|g_{m+1,m}|^2-|g_{m,m-1}|^2}{\beta} \ln (\sqrt{\beta}|t|).
\end{equation} 
The parameter 
$$
g_{m+1,m}=g(m+1)\sqrt{N-m}
$$
describes coupling between the states with $m+1$ and $m$ split molecules.

Let the system be in the state \( |\psi^0_{-}(t)\rangle \) as \( t \rightarrow -\infty \). Then, in the limit \( t \rightarrow +\infty \), the wavefunction evolves as
\[
|\psi(t)\rangle_{t \rightarrow +\infty} = \sum_{m=0}^N S_{m0}(g)\, |\psi^m_{+}(t)\rangle,
\]
where
\[
S_{m0}(g) = \sqrt{P_m}\, e^{i\Phi_m(g)}
\]
are the scattering amplitudes, with \( \Phi_m(g) \) denoting their respective phases.


The solution of the model predicts the following time-independent scattering phases:
\begin{widetext}
\begin{equation}
\Phi_{m}(g) = \frac{3m\pi}{4} + \sum_{k=1}^{m} \left\{ \arg\Gamma\left[i\left(k + \frac{(N-m)g^2}{\beta}\right)\right] - 2 \arg\Gamma\left(i \frac{k g^2}{\beta}\right) \right\}, \quad \text{for} \quad m = 1, \ldots, N,
\label{phase-degenerate}
\end{equation}
\end{widetext}
with \( \Phi_0(g) = 0 \), and where $\arg\Gamma[\ldots]$ is the phase of a complex-valued Euler's Gamma-function. 
We present Eq.~(\ref{phase-degenerate}) without derivation, but support it with numerical evidence. In Fig.~\ref{phase3degenerate}, we compare the analytical expression in Eq.~(\ref{phase-degenerate}) with the results of our numerical simulation for the case $ N = 3 $ (i.e., four interacting states), demonstrating excellent agreement.

\section{Two-channel reaction model}
\label{TC-sed}

 In order to explore a competition between different reaction channels, we will consider a version of the Tavis-Cummings model with two reaction channels:
\begin{eqnarray}
  \label{tcH}  
\nonumber H_{\rm TC}&=&\beta t \Psi^{\dagger} \Psi - \varepsilon (a_1^{\dagger} a_1+b_1^{\dagger} b_1 )+\varepsilon (a_2^{\dagger} a_2+b_2^{\dagger} b_2 )\\
&+& g\left[\Psi^{\dagger} (a_1b_1+a_2b_2)+ {\rm h.c.}\right]. 
\end{eqnarray}
Here, $\Psi$ is the annihilation operator of a bosonic molecule that can split into bosonic particles, $a$ and $b$,  also having indices $1$ and $2$, to which we will refer as to the first and second modes. Operators with different indices commute with one another.

The emerging bosons have an internal degree of freedom, for example, the direction of the momentum. This is taken into account by the difference of the particles $a$ and $b$ with the same index. Thus, if the momentum is conserved, then the decay of the  field $\Psi$ in its frame leads to production of the pairs of particles having opposite momenta but the same energy, which explains that, for instance, $a_1$ and $b_1$ have the same chemical potential $-\varepsilon$.  The particles with different indices are more distinct as they have different chemical potentials, $\mp \varepsilon$ for indices, respectively, $1$ and $2$. 

The number of particles of $a$ and $b$ type produced in each channel is the same. Therefore, it is convenient to work in the basis of states, $|n_1,n_2\rangle$, which correspond to $n_1$ molecules split via the first and $n_2$ molecules split via the second reaction channel. Such states are the eigenstates of the operators
\begin{equation}
\hat{n}_{1,2} = \frac{1}{2}\left(a_{1,2}^{\dagger}a_{1,2} +b_{1,2}^{\dagger}b_{1,2} \right).
\label{n12-def}
\end{equation}
From the particle conservation, it follows that the number of unsplit molecules in the state $|n_1,n_2\rangle$ is 
$$
n=N-n_1-n_2.
$$

In what follows, we will assume that $\varepsilon >0$ but $\varepsilon \ll g$. Our goal is to explore the effect of such $\varepsilon$  on the asymmetry of the final distribution of the atoms. We assume that initially we have only a large number $N$ of molecules. As $t\rightarrow -\infty$, this state is the ground state, however, near $t=0$, the system enters in resonances, at which molecules can dissociate.  

\vspace{0.2cm}
\paragraph*{\it Transition probabilities.} The model (\ref{tcH}) is solvable exactly, so the joint probability distribution to find $n_1$ pairs of atoms $a_1,\, b_1$ and $n_2$ pairs of the atoms $a_2$ and $b_2$ is known. According to Eq.~(\ref{P-itter}) applied to only two reaction channels, we find 
\be
P_{n_1,n_2} = P_{n_1}x^{N-n_1-n_2}(x^{N-n_1-n_2+1},x)_{n_2}. 
\label{P-joined}
\ee

\begin{figure}[t!]
\centering \includegraphics[width=0.7\columnwidth]{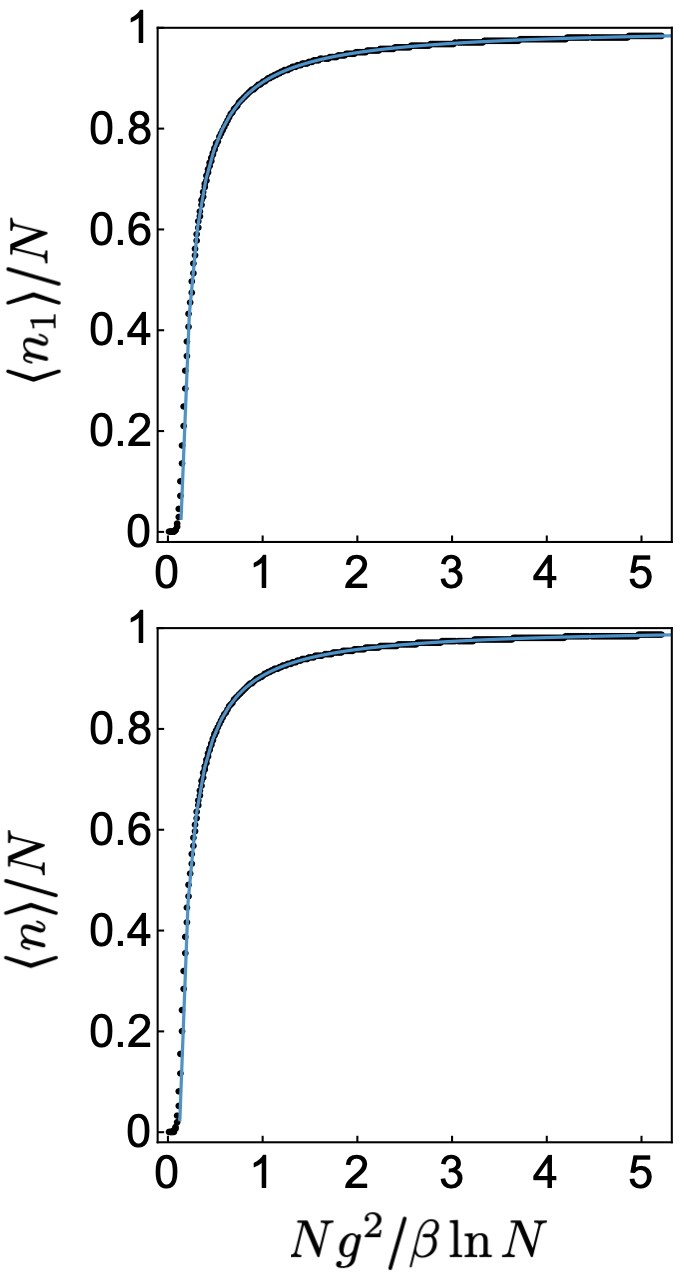}
\caption{$\langle n_1 \rangle/N$ (upper panel) and $\langle n\rangle/N$ (lower panel) as the function of $Ng^2/(\beta \ln N)$. The dots are obtained numerically using   Eq.~(\ref{P-one}) and Eq.~(\ref{Pn-fin}), respectively. The blue lines represent the theoretical predictions given by   Eq.~(\ref{nu1-av-fin}) and Eq.~(\ref{nav}), respectively. Here, $N=10^5$, $g=1$.
}
\label{plot1}
\end{figure}

From Eqs.~(\ref{P-joined}) and (\ref{P-one}), we can derive the probability of finding the total number 
$$
n=n_1+n_2
$$
of the decayed molecules. For this, we express $n_2=n-n_1$ and sum Eq.~(\ref{P-joined}) over $n_1$ taking values in the range $n_1=0,1,\ldots,n$. On the way, we also use that, from the definition (\ref{QP-def}), we have
$$
 (x^{N-n+1},x)_{n-n_1} (x^{N-n_1+1},x)_{n_1} =
(x^{N-n+1},x)_{n},
$$
and that 
$$
\sum_{n_1=0}^{n} x^{-n_1} = \frac{1-x^{n+1}}{x^{n}(1-x)}.
$$
Thus, we find, for $n=0,1,\ldots,N$,
\be
P_n\!=\! \frac{x^{2(N-n)}[1-x^{n+1}]}{1-x}(x^{N-n+1},x)_{n} .
\label{Pn-fin}
\ee
Equations~(\ref{P-one}) and (\ref{Pn-fin}) are suitable for fast numerical calculations of the averages of $n$ and $n_1$, as we demonstrate in Fig.~\ref{plot1}  for $N=10^5$. 

The asymmetry of the particle production via different channels is characterized by a parameter 
\begin{equation}
\eta \equiv \frac{\la n_1 \ra -\la n_2 \ra}{\la n \ra}=2\frac{\la n_1 \ra}{\la n \ra} - 1.
\label{eta-def}
\end{equation}
The perfect symmetry corresponds to  $\la n_2 \ra =\la n_1 \ra$, and $\eta=0$. Strong asymmetry corresponds to $\la n_2 \ra /\la n_1 \ra\ll 1$, and thus to $\eta \rightarrow 1$.

 \begin{figure}[t!]
\centering \includegraphics[width=0.73\columnwidth]{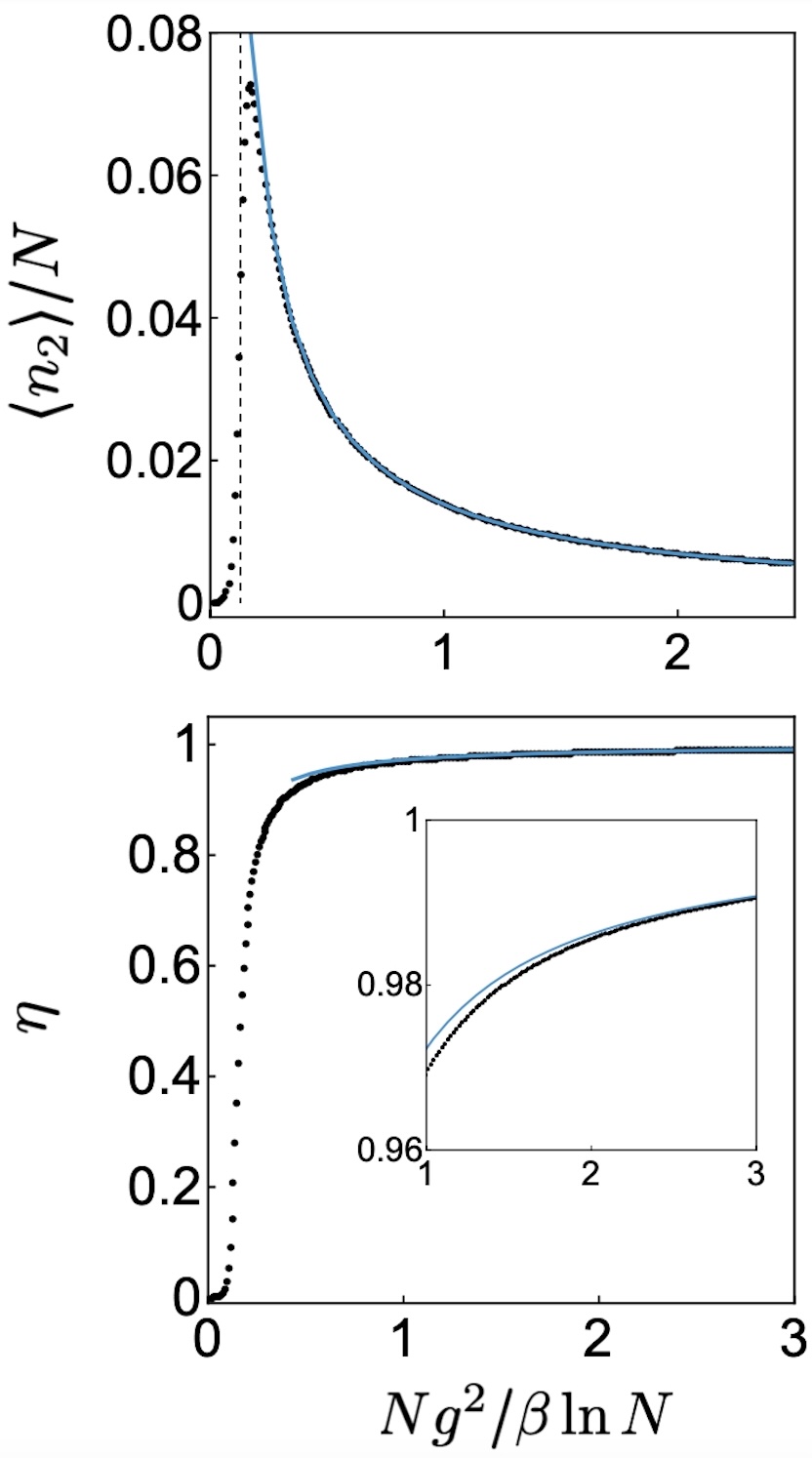}
\caption{$\langle n_2 \rangle/N$ (upper panel) and $\eta$ (lower panel) as the function of $Ng^2/(\beta \ln N)$. The dots are obtained numerically using   Eqs.~(\ref{P-one},\ref{Pn-fin})  and  Eq.~(\ref{eta-def}), respectively. The blue line represents the theoretical prediction given by    Eqs.~(\ref{n2-av}) and (\ref{eta-qa}), respectively. In the upper panel, the dashed vertical line indicates $\beta_c$ from Eq.~(\ref{betac1}). Here, $N=10^5$, $g=1$
}
\label{plot2}
\end{figure}

\vspace{0.2cm}
\paragraph*{\it Fast transition through the resonance.} We define this regime by the condition 
$$
p\equiv x^{N}\gg 1/N,
$$
which means that $\beta \gg 2\pi g^2 N /\ln N$.
We can then assume that $n_1\ll N$ for all $n_1$ that have substantially nonzero probability of occurrence, and hence,  $x^{-n_1}\approx 1$. Then, we
approximately find that $P_{n_1}$ is the geometric distribution
$$
P_{n_1}^{\rm f} \approx p(1-p)^{n_1},
$$
where ``f" stands for ``fast".
With the same accuracy, the average of $n_1$  for the rapid transition is given by
\be
\la n_1\ra^{\rm f} \approx \frac{1-p}{p}=e^{N 2\pi g^2/\beta}-1.
\label{n1-f}
\ee

Analogous calculations produce for the rapid transition
$$
P_n^{\rm f} \approx (n+1)p^{2}(1-p)^n.
$$
From this distribution, we find the average in the rapid transition regime:
\be
\la n \ra^{\rm f} \approx 2\frac{1-p}{p},
\label{nav-f}
\ee
and from (\ref{eta-def}) we find that  
\be
\eta^{\rm f} =0+O(\ln N/N).
\label{eta-f}
\ee
When the number of original molecules is macroscopic, deviations from the result (\ref{eta-f}) are negligible. Thus, even when the number of the produced particles is large, there is no significant asymmetry in the particle production. 

\vspace{0.2cm}
\paragraph*{\it Quasi-adiabatic regime.}
Let us now turn to the probability distribution for the total number of particles, $n=n_1+n_2$, in the quasi-adiabatic regime. Note that the distribution of $n_1$ coincides with the distribution of $m$ for the single channel model, and its average deviation from $N$ in the quasi-adiabatic regime is still described by Eq.~(\ref{nu1-av-fin}), where now we interpret $\nu=N-n_1$.

In the same semiclassical limit, for the two-channel reaction, we introduce the deviation 
$$
\nu_+ \equiv N-n=N-n_1-n_2,
$$
which is the number of molecules that do not split in the two-channel model.
The distribution~(\ref{Pn-fin}) corresponds then to the Euler distribution (\ref{p-euler}) for $\nu_+$ at $\alpha=x^2$, so 
\be
\la \nu_+ \ra=\frac{\ln(1-x)+\psi_x(2)}{\ln x},
\label{nav}
\ee
where we approximate 
$$
\psi_x(2) \approx \psi_1(2)=1-\gamma_E=0.422784\ldots .
$$
Hence, 
\be
\la \nu_+ \ra \approx 
\frac{
\beta}{2\pi g^2}
\left[ \ln\left(\frac{\beta}{2\pi g^2}\right)-\psi_1(2) \right].
\label{nu-av-fin}
\ee

Using that 
$$
n_2=n-n_1=\nu-\nu_+,
$$
we find the average number of the molecules that dissociate via the second reaction channel: 
\be
\la n_2 \ra =\frac{\beta \{\gamma_E+\psi_1(2)\}}{2\pi g^2}=\frac{\beta }{2\pi g^2}.
\label{n2-av}
\ee
In Fig.~\ref{plot2} we show the dependence of the average number of particle pairs in the second mode and the corresponding asymmetry parameter $\eta$ in the broad range of the sweep rates. 
Note that 
$\la n_1\ra \sim N-\la \nu \ra$, i.e., for small $\beta$ the number of the molecules dissociated via the first channel is almost the entire $N$, whereas the number of atomic pairs ending on the second mode, according to (\ref{n2-av}), decays, for $\beta \rightarrow 0$, to zero. Given that the initial number of molecules, $N$, is macroscopically large, we find that the quasi-adiabatic regime corresponds to $\eta \approx 1$. This ensures a considerable asymmetry despite the difference of the chemical potentials of the product particles, $2\varepsilon$, can be arbitrarily small. Substituting (\ref{nu1-av-fin}) and (\ref{nu-av-fin}) into (\ref{eta-def}), we determine the leading order correction to $\eta^{qa}=1$ result, for the quasi-adiabatic regime:

\be
\eta^{qa}\approx 1-\frac{2\la n_2\ra}{N}.
\label{eta-qa}
\ee

\vspace{0.2cm}
\paragraph*{\it Adiabatic regime.} 
The adiabatic limit is generally defined by the requirement 
$$
<m|dH/dt|n>/(\Delta E_{mn})^2\ll 1,
$$
where $<m|dH/dt|n>$ is the nonadiabatic characteristic coupling between the adiabatic energy levels $m$ and $n$, and $\Delta E_{mn}$ is the gap between these levels. In the Tavis-Cummings model with $\varepsilon \ll g$, one naively may expect that this corresponds to the condition $\beta/\varepsilon^2 \ll 1$.  However, the exact solution shows that both the complete molecular dissociation and the perfect asymmetry, $\eta=1$, is achieved up to exponentially suppressed corrections, $\sim e^{-f/\beta}$, when 
$$
\beta < 2\pi g^2,
$$
independently of $\varepsilon$.

\section{Phase transition during asymmetry degeneration}
\label{PT}
 Below, we illustrate that the system in the two-channel reaction model, in the large $N$-limit, passes through an additional phase transition that explains many properties of the  discussed distributions.

\subsection{Time-scale separation in $\varepsilon \rightarrow 0$ limit}
For $\beta t\gg g$, the molecular dissociation reaction is suppressed; however, as in the Demkov-Osherov model, the presence of the high energy resonance induces virtual transitions between different atomic pairs. Such transitions do not change the total number of dissociated molecules but rather redistribute the atoms between the two modes. 

Analogously to the Demkov-Osherov Model, we  first disregard the small parameter $\varepsilon$ and find that we are dealing with evolution in the subspace of states with identical state amplitudes for the same total number of atomic pairs, $n$:
\begin{equation}
|\psi_n\rangle = \frac{1}{\sqrt{n+1}}\sum_{n_1=0}^n |n_1,n-n_1\rangle, \quad n=0,1,\ldots,
\label{psi-sim}
\end{equation}
where $|n_1,n_2\rangle$ is the state with $n_1$ and $n_2$ pairs of atoms that were produced in, respectively, the first and second reaction channels. 

During time interval of order $\tau_{LZ}\sim g/\beta$, the probabilities of the states $|\psi_n\rangle$ saturate at the values given by the distribution in Eq.~(\ref{Pn-fin}). After this, the total number of the dissociated molecules is conserved, and each of the states  $|\psi_n\rangle$ evolves independently due to the virtual transitions via the states that have the same $n=n_1+n_2$, e.g., between $(n_1,n_2)$ and $(n_1+1,n_2-1)$ pairs. Using the second order perturbation theory for large $t$ in Eq.~(\ref{tcH}), we find an effective Hamiltonian that acts in phase space of the two types of atomic pairs: 
\begin{eqnarray}
\label{HeffTC}
\nonumber H_{\rm eff} &=& -\varepsilon (a_1^{\dagger} a_1+b_1^{\dagger} b_1 )+\varepsilon (a_2^{\dagger} a_2+b_2^{\dagger} b_2 ) \\
&-& \frac{g^2}{\beta t} \sum_{k,s=1,2} (a_kb_k)^{\dagger}(a_sb_s).
\end{eqnarray}
This Hamiltonian conserves the total number of bosonic pairs, so in the state space of $|n-n_2,n_2\rangle$, where $n_2=0,1,\ldots n$, it has matrix elements 

\begin{widetext}
\begin{eqnarray}
\label{mt-el1}
&&\langle n_2,n-n_2 |H_{\rm eff}|n-n_,n_2\rangle = \varepsilon (2n_2-n) -\frac{g^2[n_2^2+(n-n_2)^2]}{\beta t},\\
\nonumber &&\langle n_2+1,n-n_2-1 |H_{\rm eff}|n-n_2,n_2\rangle = -\frac{g^2[(n-n_2)(n_2+1)]}{\beta t}.
\label{Heff-mt}
\end{eqnarray}
\end{widetext}
The state (\ref{psi-sim}), which emerges after the molecular dissociation, is the ground state of this Hamiltonian within the sector with $n=n_1+n_2$ as $t\rightarrow 0_+$. The energy of this state at $t\sim \tau_{LZ}$ is 
$$
E_0(t\rightarrow 0_+)= -\frac{g^2n(n+1)}{\beta t}.
$$

 \begin{figure}[t!]
\centering \includegraphics[width=0.8\columnwidth]{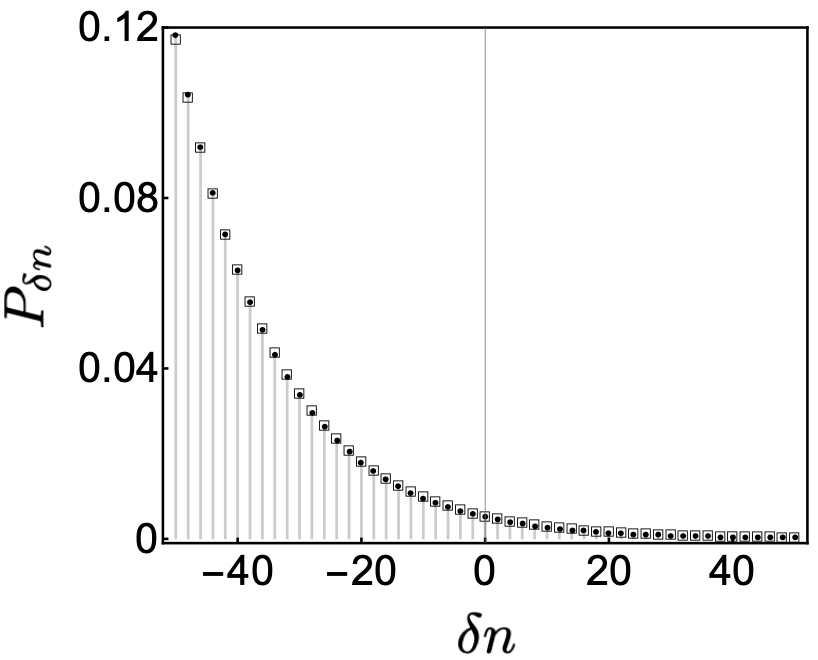}
\caption{$P_{\delta n}$ as the function of $\delta n$. Theoretical predictions (dots on vertical lines) of Eq.~(\ref{Dist}) are compared to results of a numerical solution of the time-dependent Schr\"odinger equation (empty boxes) with the Hamiltonian (\ref{HeffTC}). Here, $n=50, g=1, \beta=50, \epsilon=1$.
}
\label{plot3}
\end{figure}
The model (\ref{HeffTC}) is, in principle, a novel integrable time-dependent model, related to time-dependent Gaudin magnets \cite{Li2018,Zabalo2022,Barik2024}. 
Its relationship to the two-mode Tavis-Cummings model allows us to derive the probabilities of the eigenstates as $t\rightarrow +\infty$ using the distribution (\ref{P-joined}). Namely, the joint distribution (\ref{P-joined}) for the emerging atomic pairs in the two channel reaction 
is not a simple thermal distribution, for example,  because the ratio 
$P_{n_1,n_2}/P_{n_1+1,n_2}$ depends on $n_1,n_2$. However, if we fix the total number of the atomic pairs, $n_1+n_2=n$, then the ratio of the conditioned probabilities satisfies the detailed balance condition: 
\be
\frac{P(n_1,n_2|n)}{P(n_1+1,n_2-1|n)}=x=e^{-2\pi g^2/\beta}.
\ee
If we consider the evolution with the Hamiltonian (\ref{HeffTC}) that starts as $t\rightarrow 0_+$ with the state (\ref{psi-sim}), then as $t\rightarrow +\infty$, the probability of finding $n_2$ bosons in the 2nd mode and $n_1=n-n_2$ bosons in the 1st mode is given by 
\begin{equation}
   P_{\delta n}=\frac{1}{Z} e^{-\delta n/k_BT},
    \label{Dist}
\end{equation}
where
$$
\delta n \equiv n_2-n_1, \quad k_BT =\frac{\beta}{\pi g^2},
$$
and the normalization constant is given by 
$$
Z=\sum_{ k=-n/2}^{n/2} e^{-2k/k_BT}=\frac{\sinh \left(\frac{1+n}{k_BT} \right)}{\sinh\left(\frac{1}{k_BT} \right)}.
$$

In Fig.~\ref{plot3} we confirm the analytical solution in Eq.~(\ref{Dist}), of the model (\ref{HeffTC}), using results of our numerical simulations for $n=50$. This solution ensures that, after the total number $n$ of the atomic pairs becomes determined, the distribution within each $n$-sector thermalizes between the two reaction modes, so that the final distribution is the Gibbs distribution (\ref{Dist}).

 \begin{figure}[t!]
\centering \includegraphics[width=0.75\columnwidth]{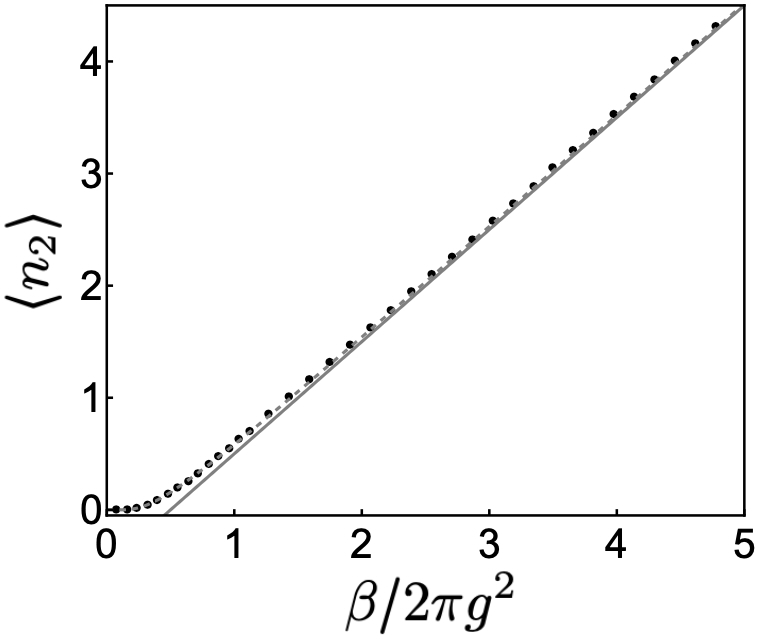}
\caption{The number of particle pairs in the higher energy mode,
$\langle n_2 \rangle$, at the end of the evolution as a function of $\beta/2\pi g^2$. The dots were obtained numerically using Eq.~(\ref{HeffTC}).   The solid line represents $\beta/2\pi g^2 -1/2$ and the dashed line represents $\coth (\pi g^2/\beta)/2-1/2$,  according to Eq.~(\ref{n2-av-esm}). $N=50$, $g=1$, $\varepsilon=1$.}
\label{n2}
\end{figure}

Apparently, as the effective temperature decreases with the sweep rate, in the adiabatic limit the system approaches the ground state with $n_2=0$. In the quasi-adiabatic regime, however, $n_2$ is the number of excitations, which is suppressed but not exponentially. Rather it follows a power-law on its own, namely, for $k_B T\ll n$ we can estimate the average $n_2$ by writing $n_2=(n+\delta n)/2$ and keeping the leading order in the $k_BT/n$ contribution:  
\begin{equation}
\langle n_2 \rangle = \sum_{k=-n/2}^{n/2} \frac{n+2k}{2} P_{2k} \approx \frac12 \coth\left(\frac{\pi g^2}{\beta}\right)-\frac12\approx \frac{\beta}{2\pi g^2}-\frac{1}{2},
\label{n2-av-esm}
\end{equation}
which agrees with Eq.~(\ref{n2-av}).

\subsection{Phase transition in two-mode competition}
Equation~(\ref{Dist}), in the quasi-adiabatic limit, exhibits a linear scaling of excitations (\ref{n2-av-esm}) versus the sweep rate $\beta$, indicating the passage through a phase transition during the time-dependent evolution. 
Let us now explore the ground state of the Hamiltonian (\ref{HeffTC}) in the mean field approximation, while treating $t$ as a constant parameter that controls the  strength of the quartic interaction.  The quadratic part in (\ref{HeffTC}) favors the mode asymmetry, with all bosons in the ground state concentrated at $n_2=0$. In contrast, the quartic interaction term supports a ground state with equal number of bosons in both modes and strong correlations between them. Consequently, it is natural to consider  the correlator between the bosons of the two modes as an order parameter, whose disappearance in the thermodynamic limit $n\rightarrow \infty$ is the signature of a phase transition between such uncorrelated and correlated phases.  

To demonstrate that this intuition is correct, we decouple the quartic interaction by introducing two mean fields: 
\begin{eqnarray}
\label{mean-eff}
\nonumber \Delta_2&\equiv& \la {a}_2^{\dagger} {a}_2 \ra=\la {b}_2^{\dagger} {b}_2 \ra, \\
\Delta_{12}&\equiv& \la {a}_2^{\dagger} {a}_1 \ra=\la {b}_2^{\dagger} {b}_1 \ra=\la {a}_2^{\dagger} {b}_1 \ra=\la {b}_2^{\dagger} {a}_1 \ra,
\end{eqnarray}
where the averaging is over the ground state of $H_{\rm eff}$.
The particle conservation requires also that 
$$
\la {a}_1^{\dagger} {a}_1 \ra = n-\Delta_2.
$$
Using such particle conservation constraints, we find that the field $\Delta_2$ contributes only to a constant term to the 
 quadratic mean field Hamiltonian. Moreover, the contributions to it of the $a$ and $b$ fields are identical:
 \begin{equation}
 H_{\rm eff}^{\rm mf} = 2\varepsilon (a_2^{\dagger} a_2-a_1^{\dagger}a_1) -\frac{4\Delta_{12} g^2}{\beta t} (a_1^{\dagger}a_2 +a_2^{\dagger}a_1),
\label{heff-mf}
 \end{equation}
 where we assume that $\Delta_{12}$ is real, which can  always be achieved by redefining phases of the bosonic operators. 

  Next, we introduce new bosonic operators, $\alpha_{1,2}$, such that
$$
a_1=\alpha_1\cos\theta +\alpha_2 \sin \theta ,
\quad a_2=-\alpha_1 \sin\theta +\alpha_2 \cos \theta ,
$$
and choose the angle $\theta$ to eliminate the mixing interactions between different bosons, so that 
$$
H_{\rm eff}^{\rm mf}=E (\alpha_2^{\dagger} \alpha_2 -\alpha_1^{\dagger} \alpha_1), \,\,\, E\equiv 2\sqrt{\varepsilon^2 +\gamma_{12}^2}, \,\,\, \gamma_{12} \equiv \frac{2\Delta_{12}g^2}{\beta t}.
$$

This requires that 
\begin{equation}
    \cos (2\theta) =\frac{\varepsilon}{\sqrt{\gamma_{12}^2+\varepsilon^2}}, \quad \sin (2\theta) =- \frac{\gamma_{12}}{\sqrt{\gamma_{12}^2+\varepsilon^2}}.
\end{equation}
In the ground state, all $n$ bosons in the system are in the mode $\alpha_1$, that is, 
$\la\alpha_1^{\dagger}\alpha_1 \ra=n$ and $\la\alpha_2^{\dagger}\alpha_2 \ra=0$. Thus, we find a self-consistent condition on $\Delta_{12}$:
$$
\Delta_{12}=-n\cos \theta \sin \theta =-n\sin (2\theta)/2,   
$$
from which we find the mean-field  
\begin{equation}
  \Delta_{12}^2=\left(\frac{n}{2}\right)^2-\frac{\varepsilon^2 (\beta t)^2}{4g^4}.
    \label{delta12}
\end{equation}
According to this estimate, at 
$$
t_c=\frac{Ng^2}{\beta \varepsilon},
$$
the mean field is zero, i.e., $\Delta_{12}(t_c)=0$. As we consider the regime with $\beta \sim N g^2/\ln N$, this phase transition occurs at time $t_c\sim \ln N/\varepsilon$.

Thus, for $t<t_c$, the mean-field ground state of the Hamiltonian (\ref{HeffTC}) corresponds to a state with nonzero $\Delta_{12}$, indicating a phase with a symmetry breaking, which corresponds to the initial choice of the phase of $\Delta_{12}$ right after the decay of the molecular condensate. In contrast, for $t>t_c$, we encounter a phase without significant cross-correlations between the different atomic modes, $1$ and $2$. 

According to the  phenomenology of critical phenomena, the quasi-adiabatic transition through this critical point should induce a power-law tail of the nonadiabatic excitations, which is in agreement with our finding in Eq.~(\ref{n2-av}). The absence of logarithmic corrections to this power law arises from the fact that here the system is transitioning from the phase with broken symmetry to the symmetric phase. For such a transition direction, a scalar field theory does not  predict a logarithmic factor in the scaling of excitations \cite{Itin-22009}.

\section{Semiclassical perspective}
\label{semicl}
\subsection{Classical limit}

The nature of the phase transitions that we have described is also revealed in the semiclassical picture, which is justified by a large number $N$ of initial molecules. For a single reaction channel, this approach was developed in \cite{Altland2009,Itin2009,Itin-22009,Sadhasivam2024}. 

Consider the nonzero matrix elements of the two-channel model Hamiltonian (\ref{tcH}) in the basis (\ref{n12-def}):
\begin{eqnarray}
\label{H_TC_matrix}
 \nonumber   && \la n_2,n_1|H|n_1,n_2 \ra = \beta t (N-n_1-n_2) + \varepsilon (n_2-n_1), \\
  \nonumber &&\la n_2,n_1-1|H|n_1,n_2 \ra  =g n_1\sqrt{N-n_1-n_2+1} , \\
  \nonumber &&\la n_2-1,n_1|H|n_1,n_2 \ra  =g  n_2\sqrt{N-n_1-n_2+1}.
\end{eqnarray}
We look for the solution of the Schr\"odinger equation in the form 
$$
|\psi \rangle = \sum_{n_1,n_2=0}^N  a_{n_1,n_2}|n_1,n_2\rangle,
$$
and introduce the amplitude generating function
\begin{align}
    u(\phi_1,\phi_2,t ) =  \sum_{n_1,n_2=0}^N a_{n_1,n_2}(t) e^{in_1\phi_1+in_2 \phi_2}.
\end{align}
Note that 
$$n_{1,2} e^{in_1\phi_1+in_2 \phi_2} = -i\partial e^{in_1\phi_1+in_2 \phi_2}/\partial \phi_{1,2},$$
so the time-dependent Schr\"odinger equation for the amplitudes can be formally written in terms of a single equation for $u(\phi_1,\phi_2,t)$ as 
\begin{equation}
\label{se-u}
i\frac{\partial }{\partial t} u(\phi_1,\phi_2,t)= \hat{H}\left(-i\frac{\partial}{\partial \phi_1}, \phi_1 ; -i\frac{\partial}{\partial \phi_2} ,\phi_2\right) u(\phi_1,\phi_2,t ),
\end{equation}
where we associate the number operators with $\hat{n}_{1,2} \equiv -i\partial/\partial \phi_{1,2}$. In the semiclassical approximation we can then associate $\phi_{1,2}$ with coordinates that are conjugate to the classical momenta $n_{1,2}$. Additionally, we can disregard the terms of order $1/N$. Thus, the classical Hamiltonian that corresponds to the 
Schr\"odinger equation (\ref{se}) with the two-channel Hamiltonian (\ref{tcH}) takes the form
\begin{eqnarray}\label{H_MF}
 \nonumber   H_{cl}(n_1,\phi_1;n_2,\phi_2) = -\beta t (n_1+n_2) +\varepsilon(n_2-n_1)+\\
    2g\sqrt{N-n_1-n_2} \left[n_1\cos(\phi_1)+n_2\cos(\phi_2)\right].
\end{eqnarray}
The classical equations of motion are given by
\begin{eqnarray}
 \nonumber  \frac{d \phi_{1}}{dt} = \frac{\partial H_{cl}}{\partial n_{1}}, \quad \frac{d n_{1}}{dt} =-\frac{\partial H_{cl}}{\partial \phi_{1}},\\
  \frac{d \phi_{2}}{dt} = \frac{\partial H_{cl}}{\partial n_{2}}, \quad \frac{d n_{2}}{dt} =-\frac{\partial H_{cl}}{\partial \phi_{2}}.
    \label{cl-em1} 
\end{eqnarray}

Equations~(\ref{cl-em1}) are integrable in the sense that we can connect the asymptotic behavior of their solutions as $t\rightarrow \pm \infty$ using the classical limit of the original solvable quantum model. Moreover, the two-mode version of the Hamiltonian (\ref{h23-c}), which commutes with the quantum two-mode Hamiltonian, also has a semiclassical limit. Let us associate the parameter $\varepsilon$ with the second time $\tau$ in (\ref{h23-c}). Analogous steps then lead to 
\begin{eqnarray}
    \label{h-cl-comm}
 \nonumber  && H'_{cl}(n_1,\phi_1;n_2,\phi_2)=
 t(n_2-n_1)-\frac{\varepsilon}{\beta}(n_1+n_2) \\
\nonumber    &&-\frac{2g}{\beta} \sqrt{N-n_1-n_2}(n_2\cos \phi_2-n_1\cos \phi_1) \\
    &&+\frac{4g^2n_1n_2}{\beta \varepsilon} \sin^2\left( \frac{\phi_1-\phi_2}{2}\right).
\end{eqnarray}

The Hamiltonians in Eqs.~(\ref{H_MF}) and (\ref{h-cl-comm}) satisfy the relations 
\begin{equation}
\frac{\partial H_{cl}}{\partial \varepsilon} = \frac{\partial H'_{cl}}{\partial t}, \quad \left \{ H_{cl},H_{cl}' \right \}=0, 
    \label{cl-consist}
\end{equation}
where 
$$
\left \{ H_{cl},H_{cl}' \right \}\equiv \sum \limits_{s=1,2} \left( \frac{\partial H_{cl}}{\partial \phi_{s}}\frac{\partial H_{cl}'}{\partial n_s}-\frac{\partial H_{cl}}{\partial n_{s}}\frac{\partial H_{cl}'}{\partial \phi_s}\right)
$$
are the standard Poisson brackets. Thus, the classical limit of the two-channel model provides an example of a realization of the MLZ-type integrability in the domain of classical physics.

\subsection{$\varphi^4$-phase transition}
The phase transition is revealed in the semiclassical Hamiltonian if we consider an evolution for a relatively small number of atomic pairs: 
$$
1\ll n_{1,2} \ll N.
$$
The stationary points then correspond to $\phi_{1,2}=\pi$. Keeping only the terms in Eq.~(\ref{H_MF}) up to the second-order in both $n_{1,2}$ and $\phi_{1,2}$, we find
\begin{eqnarray}
\nonumber H_{cl} &\approx& -(\beta t+2g\sqrt{N}) (n_1+n_2) +\varepsilon (n_2-n_1)\\
&+&\frac{g}{\sqrt{N}}(n_1+n_2)^2+g\sqrt{N}(n_1\phi_1^2+n_2\phi_2^2).
\label{hcl-3}
\end{eqnarray}

We redefine $t \rightarrow t-t_c$, where  $t_c=2g\sqrt{N}/\beta$ is the moment of reaching the first critical point. Then, we make a canonical change of variables 
\be
n_{1,2}=X_{1,2}^2, \quad \phi_{1,2}=-P_{1,2}/(2X_{1,2}),
\label{canon}
\ee
where now ${\bf X}=(X_1,X_2)$ is a vector of coordinates and ${\bf P} = (P_1,P_2)$ is the vector of momenta. We introduce new parameter combinations
$$
\gamma\equiv \frac{4g}{\sqrt{N}}, \quad m=\frac{2}{g\sqrt{N}}
$$
in order to write the Hamiltonian (\ref{hcl-3}) as
\be
H_{cl}(t,\varepsilon) = \frac{{\bf P}^2}{2m}+V({\bf X}), \quad |{\bf P}|\equiv \sqrt{P_{1}^2+P_{2}^2}.
\label{hcl-4}
\ee
where 
$$
V({\bf X}) = -\beta t |{\bf X}|^2 +\frac{\gamma}{4}|{\bf X}|^4 +\varepsilon (X_2^2-X_1^2).
$$

This is the Hamiltonian of a classical particle with two degrees of freedom, which has already been investigated in \cite{Tyagi2025}. The  potential energy for this motion is time-dependent.  For $\varepsilon = 0$, the system clearly experiences a phase transition because, for $t<0$,  the potential energy has a single minimum at $X_{1}=X_{2}=0$ but for $t>0$ the classical minimum of the potential energy is at 
$$
|{\bf X}_{\rm min}|^2 \approx 2\beta t/\gamma. 
$$
The direction of this vector, however, is determined by a subsequent dynamics in which the term with $\varepsilon$ plays a decisive role. The fact that the system passes through this critical point explains the power-law of the excitations.

The studies of the classical dynamics with the Hamiltonian (\ref{hcl-4}) predicted two types of excitations in the semiclassical limit of a scalar ${\bm \varphi}^4$ field theory: massive Higgs bosons and almost massless Goldstone bosons, whose numbers scale with $\beta$ as, respectively, $\propto \beta \ln \beta$ and $\propto \beta$. Thus, for the two-channel Tavis-Cummings model, the Higgs bosonic excitations in \cite{Tyagi2025} correspond to the molecules that do not split into atoms, with their average number given by Eq.~(\ref{nu-av-fin}). The Goldstone bosons are the atoms that emerge in the excited mode after the molecular splitting, with their average number given by Eq.~(\ref{n2-av}).

\section{Discussion}
The driven Tavis-Cummings model offers an opportunity to investigate the competition between purely coherent chemical reactions occurring among macroscopic Bose-Einstein condensates. Its exact analytical solution fully confirms the semiclassical phenomenology that was proposed in \cite{Tyagi2025} for phase transitions in well-mixed systems. In particular, the analytical nonperturbative solution of the quantum mechanical model corroborates the formulas for the excitation scaling. 

Our additional finding is that the process of the decay of an unstable vacuum is generally accomplished by two rather than one phase transitions. As in the elementary Demkov-Osherov model, the first transition is associated with the decay of the initial quantum state. This transition is governed by the dominant energy scales $g$ and $\beta^{1/2}$, and therefore is relatively fast. 
The state of the product particles  after this transition is initially a  superposition of states with, on average, equal amounts of different product types. Nevertheless, in later times, it experiences slow pseudo-thermalization and cooling down due to the interactions via the virtual molecular state. This process terminates through another quantum phase transition at time distance $\propto 1/\varepsilon$ from the first one. 

Semiclassical theory demonstrates that the two-channel reaction model belongs to the universality class of the complex ${\bm \varphi}^4$ field theory, as introduced in \cite{Tyagi2025}. Other recent studies \cite{Sadhasivam2024, suzuki} have shown that additional terms in the Tavis-Cummings model that break its integrability keep this model in the same universality class. Therefore, the predictions for the scaling and the types of the phase transitions are robust against such model generalizations. As experiments with coherent molecular condensate dissociation into bosonic atoms have recently become feasible \cite{Zhang2021}, our analytical predictions should also be  verifiable by measuring the reaction efficiency as a function of the sweep rate of the magnetic field, which drives the system through the Feshbach resonance.

\begin{acknowledgements}
This work was supported primarily by the U.S. Department of Energy, Office of Science, Office of Advanced Scientific Computing Research, through the Quantum Internet to
Accelerate Scientific Discovery Program, and in part by the
U.S. Department of Energy, Office of Science, Basic Energy
Sciences, under Award Number DE-SC0022134.
F.S.
acknowledges support from the Los Alamos National Lab-
oratory LDRD program under project number 20230049DR
and the Center for Nonlinear Studies under project number
20250614CR-NLS.
\end{acknowledgements}

\bibliographystyle{apsrev4-2}
\bibliography{ref}

\end{document}

%% file: shortcut.tex
  \newcommand{\rsym}[1]{\ensuremath{%
  \mathfrak{r}_{\mathrm{\ifx\@empty{}\else{#1}\fi}}}}

  \definecolor{cEdit}{rgb}{.17,.44,.76}

%% file: main.bbl
\begin{thebibliography}{45}%
\makeatletter
\providecommand \@ifxundefined [1]{%
 \@ifx{#1\undefined}
}%
\providecommand \@ifnum [1]{%
 \ifnum #1\expandafter \@firstoftwo
 \else \expandafter \@secondoftwo
 \fi
}%
\providecommand \@ifx [1]{%
 \ifx #1\expandafter \@firstoftwo
 \else \expandafter \@secondoftwo
 \fi
}%
\providecommand \natexlab [1]{#1}%
\providecommand \enquote  [1]{``#1''}%
\providecommand \bibnamefont  [1]{#1}%
\providecommand \bibfnamefont [1]{#1}%
\providecommand \citenamefont [1]{#1}%
\providecommand \href@noop [0]{\@secondoftwo}%
\providecommand \href [0]{\begingroup \@sanitize@url \@href}%
\providecommand \@href[1]{\@@startlink{#1}\@@href}%
\providecommand \@@href[1]{\endgroup#1\@@endlink}%
\providecommand \@sanitize@url [0]{\catcode `\\12\catcode `\$12\catcode `\&12\catcode `\#12\catcode `\^12\catcode `\_12\catcode `\%12\relax}%
\providecommand \@@startlink[1]{}%
\providecommand \@@endlink[0]{}%
\providecommand \url  [0]{\begingroup\@sanitize@url \@url }%
\providecommand \@url [1]{\endgroup\@href {#1}{\urlprefix }}%
\providecommand \urlprefix  [0]{URL }%
\providecommand \Eprint [0]{\href }%
\providecommand \doibase [0]{https://doi.org/}%
\providecommand \selectlanguage [0]{\@gobble}%
\providecommand \bibinfo  [0]{\@secondoftwo}%
\providecommand \bibfield  [0]{\@secondoftwo}%
\providecommand \translation [1]{[#1]}%
\providecommand \BibitemOpen [0]{}%
\providecommand \bibitemStop [0]{}%
\providecommand \bibitemNoStop [0]{.\EOS\space}%
\providecommand \EOS [0]{\spacefactor3000\relax}%
\providecommand \BibitemShut  [1]{\csname bibitem#1\endcsname}%
\let\auto@bib@innerbib\@empty
\bibitem [{\citenamefont {Brundobler}\ and\ \citenamefont {Elser}(1993)}]{Brundobler1993}%
  \BibitemOpen
  \bibfield  {author} {\bibinfo {author} {\bibfnamefont {S.}~\bibnamefont {Brundobler}}\ and\ \bibinfo {author} {\bibfnamefont {V.}~\bibnamefont {Elser}},\ }\href {https://doi.org/10.1088/0305-4470/26/5/037} {\bibfield  {journal} {\bibinfo  {journal} {Journal of Physics A: Mathematical and General}\ }\textbf {\bibinfo {volume} {26}},\ \bibinfo {pages} {1211} (\bibinfo {year} {1993})}\BibitemShut {NoStop}%
\bibitem [{\citenamefont {Harmin}\ and\ \citenamefont {Price}(1994)}]{Harmin1994}%
  \BibitemOpen
  \bibfield  {author} {\bibinfo {author} {\bibfnamefont {D.~A.}\ \bibnamefont {Harmin}}\ and\ \bibinfo {author} {\bibfnamefont {P.~N.}\ \bibnamefont {Price}},\ }\href {https://doi.org/10.1103/PhysRevA.49.1933} {\bibfield  {journal} {\bibinfo  {journal} {Phys. Rev. A}\ }\textbf {\bibinfo {volume} {49}},\ \bibinfo {pages} {1933} (\bibinfo {year} {1994})}\BibitemShut {NoStop}%
\bibitem [{\citenamefont {Harmin}(1997)}]{Harmin1997}%
  \BibitemOpen
  \bibfield  {author} {\bibinfo {author} {\bibfnamefont {D.~A.}\ \bibnamefont {Harmin}},\ }\href {https://doi.org/10.1103/PhysRevA.56.232} {\bibfield  {journal} {\bibinfo  {journal} {Phys. Rev. A}\ }\textbf {\bibinfo {volume} {56}},\ \bibinfo {pages} {232} (\bibinfo {year} {1997})}\BibitemShut {NoStop}%
\bibitem [{\citenamefont {Shevchenko}\ \emph {et~al.}(2010)\citenamefont {Shevchenko}, \citenamefont {Ashhab},\ and\ \citenamefont {Nori}}]{Shevchenko2010}%
  \BibitemOpen
  \bibfield  {author} {\bibinfo {author} {\bibfnamefont {S.}~\bibnamefont {Shevchenko}}, \bibinfo {author} {\bibfnamefont {S.}~\bibnamefont {Ashhab}},\ and\ \bibinfo {author} {\bibfnamefont {F.}~\bibnamefont {Nori}},\ }\href {https://doi.org/https://doi.org/10.1016/j.physrep.2010.03.002} {\bibfield  {journal} {\bibinfo  {journal} {Physics Reports}\ }\textbf {\bibinfo {volume} {492}},\ \bibinfo {pages} {1} (\bibinfo {year} {2010})}\BibitemShut {NoStop}%
\bibitem [{\citenamefont {Mi}\ \emph {et~al.}(2018)\citenamefont {Mi}, \citenamefont {Kohler},\ and\ \citenamefont {Petta}}]{Petta2018}%
  \BibitemOpen
  \bibfield  {author} {\bibinfo {author} {\bibfnamefont {X.}~\bibnamefont {Mi}}, \bibinfo {author} {\bibfnamefont {S.}~\bibnamefont {Kohler}},\ and\ \bibinfo {author} {\bibfnamefont {J.~R.}\ \bibnamefont {Petta}},\ }\href {https://doi.org/10.1103/PhysRevB.98.161404} {\bibfield  {journal} {\bibinfo  {journal} {Phys. Rev. B}\ }\textbf {\bibinfo {volume} {98}},\ \bibinfo {pages} {161404} (\bibinfo {year} {2018})}\BibitemShut {NoStop}%
\bibitem [{\citenamefont {Ginzel}\ \emph {et~al.}(2020)\citenamefont {Ginzel}, \citenamefont {Mills}, \citenamefont {Petta},\ and\ \citenamefont {Burkard}}]{Burkard2020}%
  \BibitemOpen
  \bibfield  {author} {\bibinfo {author} {\bibfnamefont {F.}~\bibnamefont {Ginzel}}, \bibinfo {author} {\bibfnamefont {A.~R.}\ \bibnamefont {Mills}}, \bibinfo {author} {\bibfnamefont {J.~R.}\ \bibnamefont {Petta}},\ and\ \bibinfo {author} {\bibfnamefont {G.}~\bibnamefont {Burkard}},\ }\href {https://doi.org/10.1103/PhysRevB.102.195418} {\bibfield  {journal} {\bibinfo  {journal} {Phys. Rev. B}\ }\textbf {\bibinfo {volume} {102}},\ \bibinfo {pages} {195418} (\bibinfo {year} {2020})}\BibitemShut {NoStop}%
\bibitem [{\citenamefont {Werther}\ \emph {et~al.}(2019)\citenamefont {Werther}, \citenamefont {Grossmann}, \citenamefont {Huang},\ and\ \citenamefont {Zhao}}]{Werther2019}%
  \BibitemOpen
  \bibfield  {author} {\bibinfo {author} {\bibfnamefont {M.}~\bibnamefont {Werther}}, \bibinfo {author} {\bibfnamefont {F.}~\bibnamefont {Grossmann}}, \bibinfo {author} {\bibfnamefont {Z.}~\bibnamefont {Huang}},\ and\ \bibinfo {author} {\bibfnamefont {Y.}~\bibnamefont {Zhao}},\ }\href {https://doi.org/10.1063/1.5096158} {\bibfield  {journal} {\bibinfo  {journal} {The Journal of Chemical Physics}\ }\textbf {\bibinfo {volume} {150}},\ \bibinfo {pages} {234109} (\bibinfo {year} {2019})},\ \Eprint {https://arxiv.org/abs/https://pubs.aip.org/aip/jcp/article-pdf/doi/10.1063/1.5096158/14120774/234109\_1\_online.pdf} {https://pubs.aip.org/aip/jcp/article-pdf/doi/10.1063/1.5096158/14120774/234109\_1\_online.pdf} \BibitemShut {NoStop}%
\bibitem [{\citenamefont {Bello}\ \emph {et~al.}(2020)\citenamefont {Bello}, \citenamefont {Kongsuwan}, \citenamefont {Donegan},\ and\ \citenamefont {Hess}}]{Bello2020}%
  \BibitemOpen
  \bibfield  {author} {\bibinfo {author} {\bibfnamefont {F.}~\bibnamefont {Bello}}, \bibinfo {author} {\bibfnamefont {N.}~\bibnamefont {Kongsuwan}}, \bibinfo {author} {\bibfnamefont {J.~F.}\ \bibnamefont {Donegan}},\ and\ \bibinfo {author} {\bibfnamefont {O.}~\bibnamefont {Hess}},\ }\href {https://doi.org/10.1021/acs.nanolett.0c01705} {\bibfield  {journal} {\bibinfo  {journal} {Nano Letters}\ }\textbf {\bibinfo {volume} {20}},\ \bibinfo {pages} {5830} (\bibinfo {year} {2020})}\BibitemShut {NoStop}%
\bibitem [{\citenamefont {Kervinen}\ \emph {et~al.}(2019)\citenamefont {Kervinen}, \citenamefont {Ram\'{\i}rez-Mu\~noz}, \citenamefont {V\"alimaa},\ and\ \citenamefont {Sillanp\"a\"a}}]{Kervinen2019}%
  \BibitemOpen
  \bibfield  {author} {\bibinfo {author} {\bibfnamefont {M.}~\bibnamefont {Kervinen}}, \bibinfo {author} {\bibfnamefont {J.~E.}\ \bibnamefont {Ram\'{\i}rez-Mu\~noz}}, \bibinfo {author} {\bibfnamefont {A.}~\bibnamefont {V\"alimaa}},\ and\ \bibinfo {author} {\bibfnamefont {M.~A.}\ \bibnamefont {Sillanp\"a\"a}},\ }\href {https://doi.org/10.1103/PhysRevLett.123.240401} {\bibfield  {journal} {\bibinfo  {journal} {Phys. Rev. Lett.}\ }\textbf {\bibinfo {volume} {123}},\ \bibinfo {pages} {240401} (\bibinfo {year} {2019})}\BibitemShut {NoStop}%
\bibitem [{\citenamefont {Sun}\ \emph {et~al.}(2012)\citenamefont {Sun}, \citenamefont {Ma}, \citenamefont {Wang},\ and\ \citenamefont {Nori}}]{Sun2012}%
  \BibitemOpen
  \bibfield  {author} {\bibinfo {author} {\bibfnamefont {Z.}~\bibnamefont {Sun}}, \bibinfo {author} {\bibfnamefont {J.}~\bibnamefont {Ma}}, \bibinfo {author} {\bibfnamefont {X.}~\bibnamefont {Wang}},\ and\ \bibinfo {author} {\bibfnamefont {F.}~\bibnamefont {Nori}},\ }\href {https://doi.org/10.1103/PhysRevA.86.012107} {\bibfield  {journal} {\bibinfo  {journal} {Phys. Rev. A}\ }\textbf {\bibinfo {volume} {86}},\ \bibinfo {pages} {012107} (\bibinfo {year} {2012})}\BibitemShut {NoStop}%
\bibitem [{\citenamefont {Wen}\ \emph {et~al.}(2020)\citenamefont {Wen}, \citenamefont {Ivakhnenko}, \citenamefont {Nakonechnyi}, \citenamefont {Suri}, \citenamefont {Lin}, \citenamefont {Lin}, \citenamefont {Chen}, \citenamefont {Shevchenko}, \citenamefont {Nori},\ and\ \citenamefont {Hoi}}]{Wen2020}%
  \BibitemOpen
  \bibfield  {author} {\bibinfo {author} {\bibfnamefont {P.~Y.}\ \bibnamefont {Wen}}, \bibinfo {author} {\bibfnamefont {O.~V.}\ \bibnamefont {Ivakhnenko}}, \bibinfo {author} {\bibfnamefont {M.~A.}\ \bibnamefont {Nakonechnyi}}, \bibinfo {author} {\bibfnamefont {B.}~\bibnamefont {Suri}}, \bibinfo {author} {\bibfnamefont {J.-J.}\ \bibnamefont {Lin}}, \bibinfo {author} {\bibfnamefont {W.-J.}\ \bibnamefont {Lin}}, \bibinfo {author} {\bibfnamefont {J.~C.}\ \bibnamefont {Chen}}, \bibinfo {author} {\bibfnamefont {S.~N.}\ \bibnamefont {Shevchenko}}, \bibinfo {author} {\bibfnamefont {F.}~\bibnamefont {Nori}},\ and\ \bibinfo {author} {\bibfnamefont {I.-C.}\ \bibnamefont {Hoi}},\ }\href {https://doi.org/10.1103/PhysRevB.102.075448} {\bibfield  {journal} {\bibinfo  {journal} {Phys. Rev. B}\ }\textbf {\bibinfo {volume} {102}},\ \bibinfo {pages} {075448} (\bibinfo {year} {2020})}\BibitemShut {NoStop}%
\bibitem [{\citenamefont {Malla}\ and\ \citenamefont {Raikh}(2018)}]{Malla2018}%
  \BibitemOpen
  \bibfield  {author} {\bibinfo {author} {\bibfnamefont {R.~K.}\ \bibnamefont {Malla}}\ and\ \bibinfo {author} {\bibfnamefont {M.~E.}\ \bibnamefont {Raikh}},\ }\href {https://doi.org/10.1103/PhysRevB.97.035428} {\bibfield  {journal} {\bibinfo  {journal} {Phys. Rev. B}\ }\textbf {\bibinfo {volume} {97}},\ \bibinfo {pages} {035428} (\bibinfo {year} {2018})}\BibitemShut {NoStop}%
\bibitem [{\citenamefont {Sinitsyn}\ \emph {et~al.}(2018)\citenamefont {Sinitsyn}, \citenamefont {Yuzbashyan}, \citenamefont {Chernyak}, \citenamefont {Patra},\ and\ \citenamefont {Sun}}]{Sinitsyn2018}%
  \BibitemOpen
  \bibfield  {author} {\bibinfo {author} {\bibfnamefont {N.~A.}\ \bibnamefont {Sinitsyn}}, \bibinfo {author} {\bibfnamefont {E.~A.}\ \bibnamefont {Yuzbashyan}}, \bibinfo {author} {\bibfnamefont {V.~Y.}\ \bibnamefont {Chernyak}}, \bibinfo {author} {\bibfnamefont {A.}~\bibnamefont {Patra}},\ and\ \bibinfo {author} {\bibfnamefont {C.}~\bibnamefont {Sun}},\ }\href {https://doi.org/10.1103/PhysRevLett.120.190402} {\bibfield  {journal} {\bibinfo  {journal} {Phys. Rev. Lett.}\ }\textbf {\bibinfo {volume} {120}},\ \bibinfo {pages} {190402} (\bibinfo {year} {2018})}\BibitemShut {NoStop}%
\bibitem [{\citenamefont {Yuzbashyan}(2018)}]{YUZBASHYAN2018323}%
  \BibitemOpen
  \bibfield  {author} {\bibinfo {author} {\bibfnamefont {E.~A.}\ \bibnamefont {Yuzbashyan}},\ }\href {https://doi.org/https://doi.org/10.1016/j.aop.2018.01.017} {\bibfield  {journal} {\bibinfo  {journal} {Annals of Physics}\ }\textbf {\bibinfo {volume} {392}},\ \bibinfo {pages} {323} (\bibinfo {year} {2018})}\BibitemShut {NoStop}%
\bibitem [{\citenamefont {Sinitsyn}\ and\ \citenamefont {Li}(2016)}]{Sinitsyn2016}%
  \BibitemOpen
  \bibfield  {author} {\bibinfo {author} {\bibfnamefont {N.~A.}\ \bibnamefont {Sinitsyn}}\ and\ \bibinfo {author} {\bibfnamefont {F.}~\bibnamefont {Li}},\ }\href {https://doi.org/10.1103/PhysRevA.93.063859} {\bibfield  {journal} {\bibinfo  {journal} {Phys. Rev. A}\ }\textbf {\bibinfo {volume} {93}},\ \bibinfo {pages} {063859} (\bibinfo {year} {2016})}\BibitemShut {NoStop}%
\bibitem [{\citenamefont {Chernyak}\ \emph {et~al.}(2020)\citenamefont {Chernyak}, \citenamefont {Li}, \citenamefont {Sun},\ and\ \citenamefont {Sinitsyn}}]{Chernyak2020}%
  \BibitemOpen
  \bibfield  {author} {\bibinfo {author} {\bibfnamefont {V.~Y.}\ \bibnamefont {Chernyak}}, \bibinfo {author} {\bibfnamefont {F.}~\bibnamefont {Li}}, \bibinfo {author} {\bibfnamefont {C.}~\bibnamefont {Sun}},\ and\ \bibinfo {author} {\bibfnamefont {N.~A.}\ \bibnamefont {Sinitsyn}},\ }\href {https://doi.org/10.1088/1751-8121/ab9464} {\bibfield  {journal} {\bibinfo  {journal} {Journal of Physics A: Mathematical and Theoretical}\ }\textbf {\bibinfo {volume} {53}},\ \bibinfo {pages} {295201} (\bibinfo {year} {2020})}\BibitemShut {NoStop}%
\bibitem [{\citenamefont {Malla}\ \emph {et~al.}(2021)\citenamefont {Malla}, \citenamefont {Chernyak},\ and\ \citenamefont {Sinitsyn}}]{Malla2021}%
  \BibitemOpen
  \bibfield  {author} {\bibinfo {author} {\bibfnamefont {R.~K.}\ \bibnamefont {Malla}}, \bibinfo {author} {\bibfnamefont {V.~Y.}\ \bibnamefont {Chernyak}},\ and\ \bibinfo {author} {\bibfnamefont {N.~A.}\ \bibnamefont {Sinitsyn}},\ }\href {https://doi.org/10.1103/PhysRevB.103.144301} {\bibfield  {journal} {\bibinfo  {journal} {Phys. Rev. B}\ }\textbf {\bibinfo {volume} {103}},\ \bibinfo {pages} {144301} (\bibinfo {year} {2021})}\BibitemShut {NoStop}%
\bibitem [{\citenamefont {Barik}\ \emph {et~al.}(2024)\citenamefont {Barik}, \citenamefont {Bakker}, \citenamefont {Gritsev},\ and\ \citenamefont {Yuzbashyan}}]{Barik2024}%
  \BibitemOpen
  \bibfield  {author} {\bibinfo {author} {\bibfnamefont {S.}~\bibnamefont {Barik}}, \bibinfo {author} {\bibfnamefont {L.}~\bibnamefont {Bakker}}, \bibinfo {author} {\bibfnamefont {V.}~\bibnamefont {Gritsev}},\ and\ \bibinfo {author} {\bibfnamefont {E.}~\bibnamefont {Yuzbashyan}},\ }\href@noop {} {\bibfield  {journal} {\bibinfo  {journal} {arXiv preprint arXiv:2409.17053}\ } (\bibinfo {year} {2024})}\BibitemShut {NoStop}%
\bibitem [{\citenamefont {Malla}\ \emph {et~al.}(2022)\citenamefont {Malla}, \citenamefont {Chernyak}, \citenamefont {Sun},\ and\ \citenamefont {Sinitsyn}}]{Malla2022}%
  \BibitemOpen
  \bibfield  {author} {\bibinfo {author} {\bibfnamefont {R.~K.}\ \bibnamefont {Malla}}, \bibinfo {author} {\bibfnamefont {V.~Y.}\ \bibnamefont {Chernyak}}, \bibinfo {author} {\bibfnamefont {C.}~\bibnamefont {Sun}},\ and\ \bibinfo {author} {\bibfnamefont {N.~A.}\ \bibnamefont {Sinitsyn}},\ }\href {https://doi.org/10.1103/PhysRevLett.129.033201} {\bibfield  {journal} {\bibinfo  {journal} {Phys. Rev. Lett.}\ }\textbf {\bibinfo {volume} {129}},\ \bibinfo {pages} {033201} (\bibinfo {year} {2022})}\BibitemShut {NoStop}%
\bibitem [{\citenamefont {Sun}\ and\ \citenamefont {Sinitsyn}(2016)}]{Sun2016}%
  \BibitemOpen
  \bibfield  {author} {\bibinfo {author} {\bibfnamefont {C.}~\bibnamefont {Sun}}\ and\ \bibinfo {author} {\bibfnamefont {N.~A.}\ \bibnamefont {Sinitsyn}},\ }\href {https://doi.org/10.1103/PhysRevA.94.033808} {\bibfield  {journal} {\bibinfo  {journal} {Phys. Rev. A}\ }\textbf {\bibinfo {volume} {94}},\ \bibinfo {pages} {033808} (\bibinfo {year} {2016})}\BibitemShut {NoStop}%
\bibitem [{\citenamefont {Sun}\ \emph {et~al.}(2019)\citenamefont {Sun}, \citenamefont {Chernyak}, \citenamefont {Piryatinski},\ and\ \citenamefont {Sinitsyn}}]{Sun2019}%
  \BibitemOpen
  \bibfield  {author} {\bibinfo {author} {\bibfnamefont {C.}~\bibnamefont {Sun}}, \bibinfo {author} {\bibfnamefont {V.~Y.}\ \bibnamefont {Chernyak}}, \bibinfo {author} {\bibfnamefont {A.}~\bibnamefont {Piryatinski}},\ and\ \bibinfo {author} {\bibfnamefont {N.~A.}\ \bibnamefont {Sinitsyn}},\ }\href {https://doi.org/10.1103/PhysRevLett.123.123605} {\bibfield  {journal} {\bibinfo  {journal} {Phys. Rev. Lett.}\ }\textbf {\bibinfo {volume} {123}},\ \bibinfo {pages} {123605} (\bibinfo {year} {2019})}\BibitemShut {NoStop}%
\bibitem [{\citenamefont {Zhang}\ \emph {et~al.}(2021)\citenamefont {Zhang}, \citenamefont {Chen}, \citenamefont {Yao},\ and\ \citenamefont {Chin}}]{Zhang2021}%
  \BibitemOpen
  \bibfield  {author} {\bibinfo {author} {\bibfnamefont {Z.}~\bibnamefont {Zhang}}, \bibinfo {author} {\bibfnamefont {L.}~\bibnamefont {Chen}}, \bibinfo {author} {\bibfnamefont {K.-X.}\ \bibnamefont {Yao}},\ and\ \bibinfo {author} {\bibfnamefont {C.}~\bibnamefont {Chin}},\ }\href {https://doi.org/10.1038/s41586-021-03443-0} {\bibfield  {journal} {\bibinfo  {journal} {Nature}\ }\textbf {\bibinfo {volume} {592}},\ \bibinfo {pages} {708} (\bibinfo {year} {2021})}\BibitemShut {NoStop}%
\bibitem [{\citenamefont {Lozanov}(2020)}]{Lozanov2020}%
  \BibitemOpen
  \bibfield  {author} {\bibinfo {author} {\bibfnamefont {K.~D.}\ \bibnamefont {Lozanov}},\ }in\ \href {https://doi.org/10.1093/oso/9780198855743.003.0012} {\emph {\bibinfo {booktitle} {Proceedings of the Les Houches Summer School: Post-Planck Cosmology}}},\ Vol.\ \bibinfo {volume} {100},\ \bibinfo {editor} {edited by\ \bibinfo {editor} {\bibfnamefont {C.}~\bibnamefont {Deffayet}}, \bibinfo {editor} {\bibfnamefont {E.}~\bibnamefont {Masso}},\ and\ \bibinfo {editor} {\bibfnamefont {G.}~\bibnamefont {Veneziano}}}\ (\bibinfo  {publisher} {Oxford University Press},\ \bibinfo {year} {2020})\ pp.\ \bibinfo {pages} {419--476},\ \Eprint {https://arxiv.org/abs/1907.04402} {arXiv:1907.04402 [astro-ph.CO]} \BibitemShut {NoStop}%
\bibitem [{\citenamefont {Kibble}(1976)}]{kibble}%
  \BibitemOpen
  \bibfield  {author} {\bibinfo {author} {\bibfnamefont {T.~W.~B.}\ \bibnamefont {Kibble}},\ }\href {https://doi.org/10.1088/0305-4470/9/8/029} {\bibfield  {journal} {\bibinfo  {journal} {Journal of Physics A: Mathematical and General}\ }\textbf {\bibinfo {volume} {9}},\ \bibinfo {pages} {1387} (\bibinfo {year} {1976})}\BibitemShut {NoStop}%
\bibitem [{\citenamefont {Zurek}(1996)}]{zurek}%
  \BibitemOpen
  \bibfield  {author} {\bibinfo {author} {\bibfnamefont {W.~H.}\ \bibnamefont {Zurek}},\ }\href {https://doi.org/10.1016/S0370-1573(96)00009-9} {\bibfield  {journal} {\bibinfo  {journal} {Physics Reports}\ }\textbf {\bibinfo {volume} {276}},\ \bibinfo {pages} {177} (\bibinfo {year} {1996})}\BibitemShut {NoStop}%
\bibitem [{\citenamefont {Damski}(2005)}]{damski}%
  \BibitemOpen
  \bibfield  {author} {\bibinfo {author} {\bibfnamefont {B.}~\bibnamefont {Damski}},\ }\href {https://doi.org/10.1103/PhysRevLett.95.035701} {\bibfield  {journal} {\bibinfo  {journal} {Phys. Rev. Lett.}\ }\textbf {\bibinfo {volume} {95}},\ \bibinfo {pages} {035701} (\bibinfo {year} {2005})}\BibitemShut {NoStop}%
\bibitem [{\citenamefont {Tyagi}\ \emph {et~al.}(2025)\citenamefont {Tyagi}, \citenamefont {Suzuki}, \citenamefont {Chernyak},\ and\ \citenamefont {Sinitsyn}}]{Tyagi2025}%
  \BibitemOpen
  \bibfield  {author} {\bibinfo {author} {\bibfnamefont {B.}~\bibnamefont {Tyagi}}, \bibinfo {author} {\bibfnamefont {F.}~\bibnamefont {Suzuki}}, \bibinfo {author} {\bibfnamefont {V.~A.}\ \bibnamefont {Chernyak}},\ and\ \bibinfo {author} {\bibfnamefont {N.~A.}\ \bibnamefont {Sinitsyn}},\ }\href {https://doi.org/10.1103/PhysRevA.111.032205} {\bibfield  {journal} {\bibinfo  {journal} {Phys. Rev. A}\ }\textbf {\bibinfo {volume} {111}},\ \bibinfo {pages} {032205} (\bibinfo {year} {2025})}\BibitemShut {NoStop}%
\bibitem [{\citenamefont {Carenza}\ and\ \citenamefont {Marsh}(2023)}]{axion}%
  \BibitemOpen
  \bibfield  {author} {\bibinfo {author} {\bibfnamefont {P.}~\bibnamefont {Carenza}}\ and\ \bibinfo {author} {\bibfnamefont {M.~D.}\ \bibnamefont {Marsh}},\ }\href {https://doi.org/10.1088/1475-7516/2023/04/021} {\bibfield  {journal} {\bibinfo  {journal} {Journal of Cosmology and Astroparticle Physics}\ }\textbf {\bibinfo {volume} {2023}}\bibinfo  {number} { (04)},\ \bibinfo {pages} {021}}\BibitemShut {NoStop}%
\bibitem [{\citenamefont {Silveira}\ \emph {et~al.}(2021)\citenamefont {Silveira}, \citenamefont {Vasconcellos}, \citenamefont {Luna},\ and\ \citenamefont {Hadjimichef}}]{CP1}%
  \BibitemOpen
\bibfield  {number} {  }\bibfield  {author} {\bibinfo {author} {\bibfnamefont {V.~M.~G.}\ \bibnamefont {Silveira}}, \bibinfo {author} {\bibfnamefont {C.~A.~Z.}\ \bibnamefont {Vasconcellos}}, \bibinfo {author} {\bibfnamefont {E.~G.~S.}\ \bibnamefont {Luna}},\ and\ \bibinfo {author} {\bibfnamefont {D.}~\bibnamefont {Hadjimichef}},\ }\href {https://doi.org/10.1007/JHEP03(2021)285} {\bibfield  {journal} {\bibinfo  {journal} {Journal of High Energy Physics}\ }\textbf {\bibinfo {volume} {2021}},\ \bibinfo {pages} {285} (\bibinfo {year} {2021})}\BibitemShut {NoStop}%
\bibitem [{\citenamefont {Yan}\ \emph {et~al.}(2021)\citenamefont {Yan}, \citenamefont {Chernyak}, \citenamefont {Zurek},\ and\ \citenamefont {Sinitsyn}}]{Yan2021}%
  \BibitemOpen
  \bibfield  {author} {\bibinfo {author} {\bibfnamefont {B.}~\bibnamefont {Yan}}, \bibinfo {author} {\bibfnamefont {V.~Y.}\ \bibnamefont {Chernyak}}, \bibinfo {author} {\bibfnamefont {W.~H.}\ \bibnamefont {Zurek}},\ and\ \bibinfo {author} {\bibfnamefont {N.~A.}\ \bibnamefont {Sinitsyn}},\ }\href {https://doi.org/10.1103/PhysRevLett.126.070602} {\bibfield  {journal} {\bibinfo  {journal} {Phys. Rev. Lett.}\ }\textbf {\bibinfo {volume} {126}},\ \bibinfo {pages} {070602} (\bibinfo {year} {2021})}\BibitemShut {NoStop}%
\bibitem [{\citenamefont {{Demkov}}\ and\ \citenamefont {{Osherov}}(1968)}]{Demkov1967}%
  \BibitemOpen
  \bibfield  {author} {\bibinfo {author} {\bibfnamefont {Y.~N.}\ \bibnamefont {{Demkov}}}\ and\ \bibinfo {author} {\bibfnamefont {V.~I.}\ \bibnamefont {{Osherov}}},\ }\href@noop {} {\bibfield  {journal} {\bibinfo  {journal} {Soviet Journal of Experimental and Theoretical Physics}\ }\textbf {\bibinfo {volume} {26}},\ \bibinfo {pages} {916} (\bibinfo {year} {1968})}\BibitemShut {NoStop}%
\bibitem [{\citenamefont {Sinitsyn}\ \emph {et~al.}(2017)\citenamefont {Sinitsyn}, \citenamefont {Lin},\ and\ \citenamefont {Chernyak}}]{Sinitsyn2017}%
  \BibitemOpen
  \bibfield  {author} {\bibinfo {author} {\bibfnamefont {N.~A.}\ \bibnamefont {Sinitsyn}}, \bibinfo {author} {\bibfnamefont {J.}~\bibnamefont {Lin}},\ and\ \bibinfo {author} {\bibfnamefont {V.~Y.}\ \bibnamefont {Chernyak}},\ }\href {https://doi.org/10.1103/PhysRevA.95.012140} {\bibfield  {journal} {\bibinfo  {journal} {Phys. Rev. A}\ }\textbf {\bibinfo {volume} {95}},\ \bibinfo {pages} {012140} (\bibinfo {year} {2017})}\BibitemShut {NoStop}%
\bibitem [{\citenamefont {Yurovsky}\ \emph {et~al.}(1999)\citenamefont {Yurovsky}, \citenamefont {Ben-Reuven}, \citenamefont {Julienne},\ and\ \citenamefont {Band}}]{Yurovsky1999}%
  \BibitemOpen
  \bibfield  {author} {\bibinfo {author} {\bibfnamefont {V.~A.}\ \bibnamefont {Yurovsky}}, \bibinfo {author} {\bibfnamefont {A.}~\bibnamefont {Ben-Reuven}}, \bibinfo {author} {\bibfnamefont {P.~S.}\ \bibnamefont {Julienne}},\ and\ \bibinfo {author} {\bibfnamefont {Y.~B.}\ \bibnamefont {Band}},\ }\href {https://doi.org/10.1088/0953-4075/32/8/306} {\bibfield  {journal} {\bibinfo  {journal} {Journal of Physics B: Atomic, Molecular and Optical Physics}\ }\textbf {\bibinfo {volume} {32}},\ \bibinfo {pages} {1845} (\bibinfo {year} {1999})}\BibitemShut {NoStop}%
\bibitem [{\citenamefont {Rangelov}\ \emph {et~al.}(2005)\citenamefont {Rangelov}, \citenamefont {Piilo},\ and\ \citenamefont {Vitanov}}]{Rangelov2005}%
  \BibitemOpen
  \bibfield  {author} {\bibinfo {author} {\bibfnamefont {A.~A.}\ \bibnamefont {Rangelov}}, \bibinfo {author} {\bibfnamefont {J.}~\bibnamefont {Piilo}},\ and\ \bibinfo {author} {\bibfnamefont {N.~V.}\ \bibnamefont {Vitanov}},\ }\href {https://doi.org/10.1103/PhysRevA.72.053404} {\bibfield  {journal} {\bibinfo  {journal} {Phys. Rev. A}\ }\textbf {\bibinfo {volume} {72}},\ \bibinfo {pages} {053404} (\bibinfo {year} {2005})}\BibitemShut {NoStop}%
\bibitem [{\citenamefont {Sinitsyn}(2014)}]{Sinitsyn2014}%
  \BibitemOpen
  \bibfield  {author} {\bibinfo {author} {\bibfnamefont {N.~A.}\ \bibnamefont {Sinitsyn}},\ }\href {https://doi.org/10.1103/PhysRevA.90.062509} {\bibfield  {journal} {\bibinfo  {journal} {Phys. Rev. A}\ }\textbf {\bibinfo {volume} {90}},\ \bibinfo {pages} {062509} (\bibinfo {year} {2014})}\BibitemShut {NoStop}%
\bibitem [{\citenamefont {Faddeev}\ and\ \citenamefont {Takhtajan}(2007)}]{Faddeev1987}%
  \BibitemOpen
  \bibfield  {author} {\bibinfo {author} {\bibfnamefont {L.}~\bibnamefont {Faddeev}}\ and\ \bibinfo {author} {\bibfnamefont {L.}~\bibnamefont {Takhtajan}},\ }\href@noop {} {\emph {\bibinfo {title} {{Hamiltonian} methods in the theory of solitons}}}\ (\bibinfo  {publisher} {Springer Science \& Business Media},\ \bibinfo {year} {2007})\BibitemShut {NoStop}%
\bibitem [{\citenamefont {Fokas}\ \emph {et~al.}(2006)\citenamefont {Fokas}, \citenamefont {Its}, \citenamefont {Kapaev},\ and\ \citenamefont {Novokshenov}}]{Fokas2006}%
  \BibitemOpen
  \bibfield  {author} {\bibinfo {author} {\bibfnamefont {A.~S.}\ \bibnamefont {Fokas}}, \bibinfo {author} {\bibfnamefont {A.~R.}\ \bibnamefont {Its}}, \bibinfo {author} {\bibfnamefont {A.~A.}\ \bibnamefont {Kapaev}},\ and\ \bibinfo {author} {\bibfnamefont {V.~Y.}\ \bibnamefont {Novokshenov}},\ }\href@noop {} {\emph {\bibinfo {title} {Painlev\'e Transcendents: The Riemann-Hilbert Approach}}},\ Mathematical Surveys and Monographs\ (\bibinfo  {publisher} {American Mathematical Society},\ \bibinfo {year} {2006})\BibitemShut {NoStop}%
\bibitem [{\citenamefont {Sun}(2025)}]{Sun2025}%
  \BibitemOpen
  \bibfield  {author} {\bibinfo {author} {\bibfnamefont {C.}~\bibnamefont {Sun}},\ }\href@noop {} {\bibfield  {journal} {\bibinfo  {journal} {arXiv preprint arXiv:2504.02576}\ } (\bibinfo {year} {2025})}\BibitemShut {NoStop}%
\bibitem [{\citenamefont {Itin}\ and\ \citenamefont {T{\"o}rm{\"a}}(2009)}]{Itin2009}%
  \BibitemOpen
  \bibfield  {author} {\bibinfo {author} {\bibfnamefont {A.}~\bibnamefont {Itin}}\ and\ \bibinfo {author} {\bibfnamefont {P.}~\bibnamefont {T{\"o}rm{\"a}}},\ }\href@noop {} {\bibfield  {journal} {\bibinfo  {journal} {arXiv preprint arXiv:0901.4778}\ } (\bibinfo {year} {2009})}\BibitemShut {NoStop}%
\bibitem [{\citenamefont {Sadhasivam}\ \emph {et~al.}(2024)\citenamefont {Sadhasivam}, \citenamefont {Suzuki}, \citenamefont {Yan},\ and\ \citenamefont {Sinitsyn}}]{Sadhasivam2024}%
  \BibitemOpen
  \bibfield  {author} {\bibinfo {author} {\bibfnamefont {V.~G.}\ \bibnamefont {Sadhasivam}}, \bibinfo {author} {\bibfnamefont {F.}~\bibnamefont {Suzuki}}, \bibinfo {author} {\bibfnamefont {B.}~\bibnamefont {Yan}},\ and\ \bibinfo {author} {\bibfnamefont {N.~A.}\ \bibnamefont {Sinitsyn}},\ }\href {https://doi.org/10.1038/s41467-024-54489-3} {\bibfield  {journal} {\bibinfo  {journal} {Nature Communications}\ }\textbf {\bibinfo {volume} {15}},\ \bibinfo {pages} {10246} (\bibinfo {year} {2024})}\BibitemShut {NoStop}%
\bibitem [{\citenamefont {Li}\ \emph {et~al.}(2018)\citenamefont {Li}, \citenamefont {Chernyak},\ and\ \citenamefont {Sinitsyn}}]{Li2018}%
  \BibitemOpen
  \bibfield  {author} {\bibinfo {author} {\bibfnamefont {F.}~\bibnamefont {Li}}, \bibinfo {author} {\bibfnamefont {V.~Y.}\ \bibnamefont {Chernyak}},\ and\ \bibinfo {author} {\bibfnamefont {N.~A.}\ \bibnamefont {Sinitsyn}},\ }\href {https://doi.org/10.1103/PhysRevLett.121.190601} {\bibfield  {journal} {\bibinfo  {journal} {Phys. Rev. Lett.}\ }\textbf {\bibinfo {volume} {121}},\ \bibinfo {pages} {190601} (\bibinfo {year} {2018})}\BibitemShut {NoStop}%
\bibitem [{\citenamefont {Zabalo}\ \emph {et~al.}(2022)\citenamefont {Zabalo}, \citenamefont {Wu}, \citenamefont {Pixley},\ and\ \citenamefont {Yuzbashyan}}]{Zabalo2022}%
  \BibitemOpen
  \bibfield  {author} {\bibinfo {author} {\bibfnamefont {A.}~\bibnamefont {Zabalo}}, \bibinfo {author} {\bibfnamefont {A.-K.}\ \bibnamefont {Wu}}, \bibinfo {author} {\bibfnamefont {J.~H.}\ \bibnamefont {Pixley}},\ and\ \bibinfo {author} {\bibfnamefont {E.~A.}\ \bibnamefont {Yuzbashyan}},\ }\href {https://doi.org/10.1103/PhysRevB.106.104513} {\bibfield  {journal} {\bibinfo  {journal} {Phys. Rev. B}\ }\textbf {\bibinfo {volume} {106}},\ \bibinfo {pages} {104513} (\bibinfo {year} {2022})}\BibitemShut {NoStop}%
\bibitem [{\citenamefont {Itin}\ and\ \citenamefont {T\"orm\"a}(2009)}]{Itin-22009}%
  \BibitemOpen
  \bibfield  {author} {\bibinfo {author} {\bibfnamefont {A.~P.}\ \bibnamefont {Itin}}\ and\ \bibinfo {author} {\bibfnamefont {P.}~\bibnamefont {T\"orm\"a}},\ }\href {https://doi.org/10.1103/PhysRevA.79.055602} {\bibfield  {journal} {\bibinfo  {journal} {Phys. Rev. A}\ }\textbf {\bibinfo {volume} {79}},\ \bibinfo {pages} {055602} (\bibinfo {year} {2009})}\BibitemShut {NoStop}%
\bibitem [{\citenamefont {Altland}\ \emph {et~al.}(2009)\citenamefont {Altland}, \citenamefont {Gurarie}, \citenamefont {Kriecherbauer},\ and\ \citenamefont {Polkovnikov}}]{Altland2009}%
  \BibitemOpen
  \bibfield  {author} {\bibinfo {author} {\bibfnamefont {A.}~\bibnamefont {Altland}}, \bibinfo {author} {\bibfnamefont {V.}~\bibnamefont {Gurarie}}, \bibinfo {author} {\bibfnamefont {T.}~\bibnamefont {Kriecherbauer}},\ and\ \bibinfo {author} {\bibfnamefont {A.}~\bibnamefont {Polkovnikov}},\ }\href {https://doi.org/10.1103/PhysRevA.79.042703} {\bibfield  {journal} {\bibinfo  {journal} {Phys. Rev. A}\ }\textbf {\bibinfo {volume} {79}},\ \bibinfo {pages} {042703} (\bibinfo {year} {2009})}\BibitemShut {NoStop}%
\bibitem [{\citenamefont {Suzuki}\ and\ \citenamefont {Zurek}(2024)}]{suzuki}%
  \BibitemOpen
  \bibfield  {author} {\bibinfo {author} {\bibfnamefont {F.}~\bibnamefont {Suzuki}}\ and\ \bibinfo {author} {\bibfnamefont {W.~H.}\ \bibnamefont {Zurek}},\ }\href {https://doi.org/10.1103/PhysRevLett.132.241601} {\bibfield  {journal} {\bibinfo  {journal} {Phys. Rev. Lett.}\ }\textbf {\bibinfo {volume} {132}},\ \bibinfo {pages} {241601} (\bibinfo {year} {2024})}\BibitemShut {NoStop}%
\end{thebibliography}%
